\documentstyle[12pt]{article}
\newcommand{\R}{\rm I \mkern -3mu R}
\newcommand{\N}{\rm I \mkern -3muN}
\newcommand{\op}{{\cal P}^{\uparrow}_{+}}
\newcommand{\up}{\widetilde{{\cal P}^{\uparrow}_{+}}}
\newcommand{\ul}{\widetilde{{\cal L}^{\uparrow}_{+}}}
\newcommand{\ol}{{\cal L}^{\uparrow}_{+}}
\newcommand{\xm}{X_{m}^{\eta}}
\newcommand{\xmp}{X_{m}^{+}}
\newcommand{\xf}{X_{0}^{\eta}}

\newcommand{\masp}{d\alpha^{+}_{m}}
\newcommand{\mas}{d\alpha^{\eta}_{m}}
\newcommand{\y}{Y_{m}}
\newcommand{\foc}{{\cal F}({\cal H})}
\begin{document}
\thispagestyle{empty}
\begin{flushright}
IFA-FT-394-1994, April
\end{flushright}
\bigskip\bigskip\begin{center}
{\bf \Huge{FREE FIELDS FOR ANY SPIN IN 1+2 DIMENSIONS}}
\end{center}
\vskip 1.0truecm
\centerline{\bf
D. R. Grigore\footnote{e-mail: grigore@roifa.bitnet}}
\vskip5mm
\centerline{Dept. Theor. Phys., Inst. Atomic Phys.,}
\centerline{Bucharest-M\u agurele, P. O. Box MG 6, ROM\^ANIA}
\vskip 2cm
\bigskip \nopagebreak \begin{abstract}
\noindent
We construct free fields of arbitrary spin in 1+2 dimensions
i.e. free fields for which the one-particle Hilbert space
carries a projective isometric irreducible representation of the
Poincar\'e group in 1+2 dimensions. We analyse in detail these
representations in the fiber bundle formalism and afterwards we apply Weinberg
procedure to construct the free fields. Some comments concerning
axiomatic field theory in 1+2 dimensions are also made.
\end{abstract}

\newpage\setcounter{page}1

\section{Introduction}

Physics in 1+2 dimensions is today a field  of intensive
research. As it is well known, relativistic physics is based on
the concept of quantum relativistic field. It would be desirable
to be able to extend the usual Wightman framework [1-3] to 1+2
dimensions.

To be able to formulate a quantum relativistic theory one must
first settle the question of relativistic invariance. Let us
denote by ${\cal P}$ the Poincar\'e group in 1+d dimensions. One usually
suppose that ${\cal P}$ is a group of symmetries of the system.
According to the general principles of quantum mechanics,
one has a projective isometric continous representation $U$ of
${\cal P}$ acting in the Hilbert space ${\cal H}$ of the system. From $U$
we extract a projective unitary continous representation of the
proper orthochronous Poincar\'e group $\op$ and another three
operators representing the space, time and space-time
inversions. One can justify on physical grounds that space
inversion is unitary and time and space-time inversions must be
antiunitary [4] (see lemma 9.9). Next, some cohomology results
show that, for $d \geq 2$, the projective unitary representations
$U$ of $\op$ follows from a true unitary continous representations
$W$ of the universal covering group $\up$. This group is
constructed as a semi-direct product between $\ul$ (= the
universal covering group of the Lorentz group in 1+d dimensions)
and $T(1+d)$ (= the translation group in 1+d dimensions, usually
identified with $\R^{1+d}$ considered as an additive group). We
remind that for $d=3$,
$\ul \simeq SL(2,{\bf C})$
and for $d=2$,~$\ul$ has a more complicated structure and it is
explicitely described in [5,6].

Now, a quantum field is a set of operator-valued distributions
$\phi_{n}(x)$
(we use for simplicity formal notations) such that, besides some
technical assumptions (regarding the existence and the ciclicity
of a vacuum state
$\Phi_{0}$,
the existence of a dense domain left invariant by the
representation $U$ as well as by
$\phi_{n}(x)$
and
$\phi_{n}(x)^{*}$)
one has:

(a) the relativistic covariance law:
$$
U_{h,a} \phi_{m}(x) U_{h,a}^{-1} = \sum_{n} T_{mn}(h^{-1})
\phi_{n}(\delta (h)\cdot x+a)\eqno(1.1)
$$
where:
$h \in \ul,~\delta:\ul \rightarrow \ol$
is the covering map, $T$ is a representation of $\ul$;

(b) the microscopic causality:
$$
[\phi_{m}(x),\phi_{n}(y)]_{\pm} =
[\phi_{m}(x),\phi_{n}(y)^{*}]_{\pm} = 0\eqno(1.2)
$$
for $x-y$ spacelike, and one of the possibilities $\pm$. We
leave for the moment unspecified the range of the indices $m,n$.

Now, an important point with respect to the axioms above (called
Wightman axioms) is to verify their independence and
compatibility. In other words, one must exhibit non-trivial
models verifying these axioms. In 1+3 dimensions one has the
so-called free fields. Namely, the Hilbert space is of the Fock
type and the fields are some special combinations of the
creation and annihilation operators (see [1], ch 3-2). Moreover,
the restriction of the representation $U$ to the one-particle
space induces a projective unitary irreducible representation of
the group $\op$ for $d=3$.

It is noteworthy the fact that in 1+3 dimensions one can find
for every projective isometric irreducible representation of the
Poincar\'e group, characterized by positive mass and arbitrary
spin, a free field such that the one-particle states carry
exactly such a representation in the sense above. Moreover, this
can be achieved by taking in (1.1) only some irreducible finite
dimensional representations of $SL(2,{\bf C})$. Another
important observation is the fact that if one wants to include
the inversions also, one has to accept a certain~"doubling" of the
number of fields. This comes from the corresponding~"doubling"
appearing in the study of the representations of the Poincar\'e
group including inversions for $d=3$. Similar assertions are
also valid for the zero-mass case.

The problem we want to investigate in this paper is to what extent
we can implement this program in 1+2 dimensions. We mention in
connection with this problem a recently published paper
[7] where however, instead of $\ul$ one considers only the twofold
cover $SL(2,\R)$ and takes the representation $T$ in (1.1) to be
finite dimensional. If the first objection can be easily dealt
with, the assumption that $T$ is finite dimensional is more
serious because one can obtain only free fields of integer or
half-integer spin. However, it is known that in 1+2 dimensions
the spin can take any real value [8], so one would want a system
of axioms which admits solutions in the form of free fields of
any spin. We will show that this can be achieved if one takes in
(1.1) $T$ to be infinite dimensional; more precisely $T$ must be
a direct sum of some representations from the complementary
series [8] of $\ul$ and the one-particle subspace will carry a
projective isometric irreducible representation of the whole
group ${\cal P}$.

In section 2 we will present briefly the projective irreducible
representations of $\op$ corresponding to mass greater or equal to
zero following [8]. We will give these representations mainly in
the fiber bundle formalism [4]. Next, we will study in section 3
the projective isometric irreducible representations of ${\cal P}$,
using the method developped in [9] and used in the 1+3 dimensional case.
In section 4 we will give Weinberg recipe for the construction
of a free field when the one-particle space is known [10-13] and
we will show that an obstruction appears if we consider only
projective unitary irreducible representations of $\up$. This
obstruction can be easily circumvented if we use irreducible
representations including inversions. In the last section we
will make some comments concerning PCT, spin and statistics and
all that.

\section{Unitary Irreducible representations of $\up$ in the Fiber
Bundle Formalism}

\subsection*{A Induced Representations}

Suppose
$H \times_{t} A$
is a semi-direct product of the locally compact groups
$H$
and
$A$
which verify the second axiom of countability,
$A$ is Abelian and
$
t: H \rightarrow Aut(A)
$
is a group homomorphism. To classify all the unitary irreducible
representations of
$H \times_{t} A$
one goes through the following procedure [4]:

(a) Denote by
$\hat{A}$
the dual of
$A$
and by
$(h,\omega) \rightarrow h\cdot \omega$
the action of $H$ on
$\hat{A}$
given by:
$$
(h\cdot \omega)(a) \equiv \omega(t_{h^{-1}}(a)).\eqno(2.1)
$$

(b) One computes all the $H$-orbits in
$\hat{A}$.
We suppose there exists a Borel cross section
$Z \subset \hat{A}$
intersecting once every $H$-orbit.

(c) For
$\forall \omega \in Z$
one computes the "little group":
$$
H_{\omega} \equiv \{ h \in H \vert h\cdot\omega = \omega \}.\eqno(2.2)
$$

(d) One tries to find out a complete list of all the unitary
irreducible representations of
$H_{\omega}, \forall \omega \in Z$.

(e) Let
${\cal O} \subset \hat{A}$
be a $H$-orbit in
$\hat{A},~\omega_{0} \equiv {\cal O} \cap Z$
and
$\pi$
a unitary irreducible representation of
$H_{\omega}$
acting in the (complex) Hilbert space
${\cal K}$.
For every such
$\pi$
one associates an
$(H,{\cal O},M({\cal K}))$-cocyle
$\phi^{\pi}$.
Here
$M({\cal K})$
is the group of unitary operators in the Hilbert space {\cal K}.

(f) The induction procedure associates to every couple
$({\cal O},\pi)$
as above a unitary irreducible representation
$W^{(O,\pi)}$
of
$H \times_{t} A$
acting in
${\cal H} \equiv L^{2}({\cal O},d\alpha,{\cal K})$
($\alpha$ is a $H$-quasi-invariant measure on ${\cal O}$) as follows:
$$
\left( W^{({\cal O},\pi)}_{h,a} f\right)(\omega) = \omega(a)
(r_{h}(h^{-1}\cdot\omega))^{1/2}
\phi^{\pi}(h,h^{-1}\cdot\omega) f(h^{-1}\cdot\omega)\eqno(2.3)
$$
(where
$r_{h}(\cdot)$
is a version of the Radon-Nycodym derivative
${d\alpha^{h^{-1}} \over d\alpha}$).

According to Mackey theorem every unitary representation of
$H \times_{t} A$
is unitary equivalent to a representation of the form
$W^{({\cal O},\pi)}$.
Moreover
$W^{({\cal O},\pi)}$
is unitary equivalent to
$W^{({\cal O'},\pi')}$
{\it iff} the orbits
${\cal O}$
and
${\cal O}'$
coincide and the representations
$\pi$
and
$\pi'$
are unitary equivalent. As regards irreducibility,
$W^{({\cal O},\pi)}$
is irreducible {\it iff}
$\pi$
is irreducible. The realization of
$W^{({\cal O},\pi)}$
in the form (2.3) is known as the canonical formalism.

{\bf Remark 1}: Note that this result does not suppose that $H$
is connected. This remark will be extremely useful in studying
the representations of the Poincar\'e group including inversions.

\subsection*{B Representations in Fiber Bundles}

The basic construction is the following. Let $X$ and $B$ be both
standard Borel $H$-spaces with $X$ transitive and let
$\pi: B \rightarrow X$
be a surjective Borel equivariant map. Explicitely, if
$(h,x) \mapsto h\cdot x$
and
$(h,b) \mapsto D(h)\cdot b$
are the actions of $H$ on $X$ and $B$ respectively, we require that:
$$
\pi(D(h)\cdot b) = h\cdot \pi(b).\eqno(2.4)
$$

We say that
$(X,B,G,\pi)$
(shortly $B$) is a Hilbert space bundle if for any
$x \in X$,
the fiber:
$$
B_{x} \equiv \pi^{-1}(\{x\})\eqno(2.5)
$$
is a (separable) Hilbert space with the natural Borel structure
induced by $B$ and such that:
$$
B_{x} \ni b \mapsto D(h)\cdot b \in B_{h\cdot x}
$$
is a unitary isomorphism of
$B_{x}$
onto
$B_{h\cdot x}$.
If we denote by
$<\cdot,\cdot>_{x}$
and
$\vert\cdot\vert_{x}$
the scalar product and respectively the norm on
$B_{x}$
this means that:
$$
<b,b'>_{x} = <D(h)\cdot b,D(h)\cdot b'>_{h\cdot x}\eqno(2.6)
$$
($\forall b,b' \in B_{x}$).

In particular, if
$H_{x}$
is the stability subgroup at $x$,
$D(h)$
leaves
$B_{x}$
invariant for any
$h \in H_{x}$
and
$h \mapsto D(h)$
defines a unitary representation
$D^{0}$
of
$H_{x}$
in
$B_{x}$.

A section of $B$ is any Borel map
$f: X \rightarrow B$
such that
$f(x) \in B_{x},~~\forall x \in X$.
Then one shows that
$x \mapsto \vert f(x)\vert^{2}_{x}$
is a Borel function. Let us suppose now that $\alpha$ is an
invariant $\sigma$-finite measure on $X$. Then we identify two
sections which are identical almost everywhere and define by
${\cal V}$ this factor space. Let now:
$$
{\cal H} = \{ f \in {\cal V}\vert \int\vert f(x)\vert_{x}^{2}
d\alpha (x) < \infty\}.\eqno(2.7)
$$

Then ${\cal H}$ is a Hilbert space with the scalar product:
$$
<f,g>~\equiv~\int_{X} <f(x),g(x)>_{x} d\alpha (x).\eqno(2.8)
$$

Now let us take $X$ to be a $H$-orbit ${\cal O}$ and
$H\times_{t} A$
as in subsection A. For any
$(h,a) \in H\times_{t} A$
we define
$W_{h,a}: {\cal H} \rightarrow {\cal H}$
by:
$$
(W_{h,a}f)(\omega) = \omega(a) D(h)\cdot f(h^{-1}\cdot \omega).\eqno(2.9)
$$

Then $W$ is a unitary representation of
$H\times_{t} A$
in the Hilbert space ${\cal H}$. Moreover, $W$ is unitary
equivalent to
$W^{{\cal O},D^{0}}$
where
$D^{0}$
is the representation of
$H_{\omega}$
defined above.

The main advantage of the fiber bundle realization is the
simplicity of the representation given by (2.9). In particular,
we do not need some cocycle associated to
$D^{0}$.

\subsection*{C Some Unitary Irreducible Representation of $\ul$}

We remind that the universal covering group
$H = \ul$
can be identified with
$\R \times D$
where
$$
D = \{ u \in {\bf C}\vert~~\vert u\vert < 1\}\eqno(2.10)
$$
with the composition law:
$$
(\phi_{1},u_{1})\cdot (\phi_{2},u_{2}) \equiv \left( \phi_{1}+\phi_{2}
+{1\over 2i} ln{1+e^{-2i\phi_{2}}u_{1}\bar{u_{2}} \over
1+e^{2i\phi_{2}}u_{2}\bar{u_{1}}}, {u_{1}+e^{2i\phi_{2}}u_{2} \over
e^{2i\phi_{2}}+u_{1}\bar{u_{2}}} \right).\eqno(2.11)
$$
and the logarithm is taken with
$Arg \in (-\pi/2,\pi/2)$.
As usual, the covering map is denoted by $\delta$; we do not
need the explicit expression here (see [8]).

We have anticipated in the introduction that we will need the
explicit expressions for unitary irreducible representations of
$H$ belonging to the so-called discrete series.

We denote by $F$ the vector space of analytic functions in the domain
$\vert z\vert < 1$;
$F$ becomes a pre-Hilbert space with respect to the scalar product:
$$
<f,g>_{F}~\equiv~{2l\over \pi} \int_{\vert z\vert < 1}
(1 - \vert z\vert^{2})^{2(l-1)}
\overline{f(z)} g(z) d\sigma
$$
Here
$l \in \R_{+}$
and
$\sigma$
is the surface measure on the unit disk. We denote by ${\cal K}$
the completion of $F$ with respect to the associated norm
$\vert\cdot\vert_{F}$. In ${\cal K}$ one has the following
representations
$D^{(l,\pm)}$
of $H$: they are defined on $F$ by:
$$
(D^{(l,+)}(\phi,u)f)(z) = e^{-2il\phi} (1 + e^{-2i\phi}\bar{u} z)^{-2l} (1
- \vert u\vert^{2})^{l} f\left( {z+e^{2i\phi} u\over
e^{2i\phi}+\bar{u} z}\right)\eqno(2.12)
$$
and respectively
$$
(D^{(l,-)}(\phi,u)f)(z) = e^{2il\phi} (1 + e^{2i\phi}u z)^{-2l} (1
- \vert u\vert^{2})^{l} f\left( {z+e^{-2i\phi} \bar{u}\over
e^{-2i\phi}+uz}\right)\eqno(2.13)
$$

One extends
$D^{(l,\pm)}$
to ${\cal K}$ by continuity.

Let us define the operator
$C: {\cal K} \rightarrow {\cal K}$
by:
$$
(Cf)(z)~\equiv~\overline{f(\bar{z})}.\eqno(2.14)
$$

Then one has immediately the following relations:
$$
C^{2} = id\eqno(2.15)
$$
$$
<Cf,Cg>_{F}~=~\overline{<f,g>_{F}}\eqno(2.16)
$$
and moreover:
$$
CD^{(l,\epsilon)}(h)C = D^{(l,-\epsilon)}(h),~~(\forall h \in H).
\eqno(2.17)
$$

We need now a clever choice for the generators of the Lie
algebra
$Lie(H) \equiv T_{e_{H}}(H)$
($e_{H}$
is the neutral element of $H$). The local coordinates in $\ul$ being
$\phi, u_{1}, u_{2}~~(u = u_{1} + u_{2})$
we define:
$$
L_{0} = {1\over 2} {\partial \over \partial \phi},~~
L_{1} = - {1\over 2} {\partial \over \partial u_{2}},~~
L_{2} = {1\over 2} {\partial \over \partial u_{1}}.\eqno(2.18)
$$

Then one can identify
$Lie(H) \equiv \R^{3}$
according to:
$$
\R^{3} \ni X \mapsto L_{X} \equiv X^{0} L_{0} + X^{1} L_{1} +
X^{2} L_{2} \in Lie(H)\eqno(2.19)
$$
and the adjoint action becomes extremely simple:
$$
Ad_{h}~L_{X} = L_{\delta(h)\cdot X}.\eqno(2.20)
$$

Note that the basis (2.18) differs from the choice from [8] and
was made especially to the purpose of obtaining such a
convenient adjoint action.

The reason for all this is the following one. Let us define the
infinitesimal generators of the representation
$D^{(l,\epsilon)}$
via Stone-von-Neumann theorem, with the following convention:
$$
exp\left( -itH_{X}^{(\epsilon)}\right) = D^{(l,\epsilon)}\left(
exptL_{X}\right).\eqno(2.21)
$$

Then (2.20) implies that on a suitable G\aa rding domain we have:
$$
D^{(l,\epsilon)}(h) H^{(\epsilon)}_{X}  D^{(l,\epsilon)}(h^{-1})
= H^{(\epsilon)}_{\delta(h)\cdot X}\eqno(2.22)
$$
($\forall h \in H,~\forall X
\in \R^{3}$).

Finally, we note that a basis in ${\cal K}$ is of the form:
$$
g_{m}(z) = \gamma_{m} z^{m}~~(\forall m \in {\bf Z})\eqno(2.23)
$$
with some suitable chosen coefficients
$\gamma_{m} \in \R_{+}$.
By
$D^{(l,\epsilon)}_{kk'}(h)$
we denote the corresponding matrix elements.

\subsection*{D The Unitary Irreducible Representations of $\up$ in
the Fiber Bundle Formalism}

The group $\up$ can be realized as the semidirect product
$H\times_{\delta} T(3)$.
According to [8] we have four type of orbits.

(I) $\xm~~(m \in \R_{+},~\eta = \pm)$

Then
$$
H_{\eta me_{0}} = \{(\phi,0)\vert \phi \in \R\}
$$
with the unitary irreducible representations
$\pi^{(s)}~~(s \in \R)$
acting in {\bf C} according to:
$$
\pi^{(s)}(\phi,0) = exp(is\phi).\eqno(2.24)
$$

If
$\varphi^{(s)}$
is the corresponding cocycle, then one can realize the induced
representations
$W^{m,\eta,s} \equiv W^{\xm,\pi^{(s)}}$
according to (2.3). Let us exhibit
$W^{m,+,s}$
in the fiber bundle formalism. We first consider the case
$s \in \R_{+}$.

In the framework of subsection B we take

- $$
B = B^{+,s}_{m} \equiv \left\{(p,f) \vert p \in \xmp, f \in F,
(p\cdot H^{(-)}+sm/2)f = 0\right\}.\eqno(2.25)
$$

Here
$H^{(-)}_{X}$
have been defined by (2.21) and
$p\cdot H^{(\epsilon)}$
denotes the usual Minkowski scalar product.

- $$
X = \xmp
$$

- $$
\pi(p,f) = p
$$
(the canonical projection on the first entry)

- $$
<\cdot,\cdot>_{p}~=~<\cdot,\cdot>_{F}
$$

- the action of $H$ on $B$ is:
$$
D(h)\cdot (p,f) \equiv (\delta(h) \cdot p,D^{(s/2,-)}(h)\cdot
f).\eqno(2.26)
$$

The only non-trivial thing to check is that
$D(h)$
is well defined i.e. it maps the fiber
$B^{+,s}_{m}(p)$
onto the fiber
$B^{+,s}_{m}(\delta(h)\cdot p)$.
This follows immediately from (2.22).

We now apply the scheme from section B and obtain an unitary
representation of $\up$ in the Hilbert space:
$$
{\cal H}^{+,s}_{m} \equiv \left\{ \varphi: \xm \rightarrow F \vert
\varphi~is~Borel,~\varphi(p) \in B^{+,s}_{m}(p),~\int_{\xmp}
<\varphi(p), \varphi(p)>_{F} \masp <\infty \right\}\eqno(2.27)
$$
according to:
$$
\left( W_{h,a}\varphi\right)(p) = exp(i a\cdot p) D^{(s/2,-)}(h)\cdot
\varphi(\delta(h^{-1})\cdot p).\eqno(2.28)
$$

To match this representation with
$W^{m,+,s}$
we only have to compute the restriction of
$D(h)$
above to a suitable fiber, say
$B^{+,s}_{m}(me_{0})$.
According to (2.25) the Dirac-like equation determining this
fiber is:
$$
\left( H^{(-)}_{0} + s/2\right) f = 0
$$

{}From (2.18) and (2.21) one easily determines
$H^{(-)}$
and shows that in fact this fiber is one-dimensional and
generated by
$g_{0}$
(see (2.23)) i.e.
$B^{+,s}_{m}(me_{0}) = Span(g_{0}).$
In particular all the fibers of $B$ are one-dimensional. Now it
is clear from (2.26) that:
$$
D(\phi,0)\cdot g_{0} = exp(is\phi) g_{0}
$$
so, the restriction of $D$ to
$H_{me_{0}}$
coincides with
$\pi^{(s)}$
from (2.24). This shows that (2.28) gives indeed
$W^{m,+,s}$.
The case
$s \in \R_{-}$
is solved considering
$D^{(s/2,+)}$
instead of
$D^{(s/2,-)}$.

Let us remark that Dirac-like equations of the type appearing in
(2.25) have been introduced previously in [14].

(II) $\xf~(\eta = \pm)$

Then
$(e_{+} = e_{0} + e_{1})$:

$$
H_{\eta e_{+}} = \left\{ \left( {1 \over 2i} ln {1-ib \over 1+ib}
+n\pi, {ib \over 1-ib} \right) \vert n \in Z, b \in R\right\}
$$
with the representations
$\pi^{(s,t)}~~(s \in \R(mod~ 2),~t \in \R)$
acting in {\bf C} according to:
$$
\pi^{(s,t)}\left( {1 \over 2i} ln {1-ib \over 1+ib}
+n\pi, {ib \over 1-ib} \right) \equiv e^{\pi isn} e^{itb}.\eqno(2.29)
$$

If
$\varphi^{(s,t)}$
is the corresponding cocycle then one can use (2.3) to determine
$W^{\eta,s,t} \equiv W^{\xf,\pi^{(s,t)}}$.
It is tempting to make formally
$m \rightarrow 0$
in the formulae from (I) and see what representation one
obtains. So, we keep (2.25)-(2.28) with
$m = 0$.
Now we select the fiber
$B^{+,s}_{0}(e_{+})$
determined by the equation:
$$
\left( H^{(-)}_{0} - H^{(-)}_{1}\right) f = 0.
$$

Again, the fiber is one-dimensional:
$B^{+,s}_{0}(e_{+}) = Span(f_{0})$,
where:
$$
f_{0}(z) = {1\over (z+1)^{s}}.\eqno(2.30)
$$

Because we also have:
$$
D\left( {1 \over 2i} ln {1-ib \over 1+ib} +n\pi,
{ib \over 1-ib} \right) f_{0} =  e^{\pi isn} f_{0}
$$
i.e. we have obtained
$\pi^{(s,0)}$
from (2.29), it follows that the limit
$m \rightarrow 0$
in (2.25)-(2.28) gives us the representation
$W^{+,s,0}$.
It is not clear if
$W^{+,s,t}$
for
$t \not= 0$
can be obtained in the fiber bundle formalism. So, when we will talk
about free fields of zero-mass we will implicitely understand
that the one -particle spaces corresponds to
$t = 0$.
Again, the case
$s \in \R_{-}$
is taken care off using
$D^{(s/2,+)}$
instead of
$D^{(s/2,-)}.$

(III) $\y$

Then
$$
H_{me_{2}} = \left\{ \left( \pi n,{a-1\over a+1}\right) \vert n
\in {\bf Z}, a \in R_{+} \right\}
$$
with the representations
$\pi^{',s,t}~~(s \in R(mod~ 2),~t \in R)$
acting in {\bf C} according to:
$$
\pi^{',s,t}\left( \pi n,{a-1\over a+1}\right) \equiv
e^{\pi isn} a^{t}.\eqno(2.31)
$$

If
$\phi^{',s,t}$
is the corresponding cocycle, then we can use (2.3) to determine
$W^{',s,t} \equiv W^{\y,\pi^{',s,t}}$.
One may try to find out a natural way of modifying the formalism
from (I). It is clear that a good guess is
$m \rightarrow im$
i.e. one considers for
$s \in \R_{+}$:
$$
B = B^{s}_{im} \equiv \left\{(p,f) \vert p \in \y, f \in F,
(p\cdot H^{(-)}+ism/2)f = 0\right\}.\eqno(2.32)
$$
and leaves (2.26) unchanged. Then the Hilbert space of the
induced representation must be:
$$
{\cal H}^{s}_{im} \equiv \left\{ \varphi: \y \rightarrow \vert
\varphi~is~Borel,~\varphi(p) \in B^{s}_{im}(p),~\int_{\y}
<\varphi(p), \varphi(p)>_{F} d\beta_{m}(p) <\infty \right\}\eqno(2.33)
$$
and we keep unchanged (2.28). To find out what representation we
have obtained we proceed as before. The fiber
$B^{s}_{im}(me_{2})$
is determined by the equation:
$$
\left( H^{(-)}_{2} -is/2\right) f = 0
$$
and is explicitely given by
$Span(f_{0}')$
where:
$$
f_{0}'(z) = {1\over (z-1)^{s}}.\eqno(2.34)
$$

Because:
$$
D\left( \pi n,{a-1\over a+1}\right) f_{0}' = e^{\pi isn} a^{s/2}
f_{0}'
$$
it follows that (2.28) gives in this case the representation
$W^{m,s,s/2}$.
For
$s \in \R_{-}$
one must use
$D^{(s/2,+)}$.
Again, it is not clear if one can obtain
$W^{m,s,t}$
with
$t \not= s/2$
in the fiber bundle formalism.

(IV) $X_{00}$. This case will not be considered here.

\section{Projective Isometric Irreducible Representations of ${\cal P}$}
\subsection*{A Invariance to Inversions in 1+2 Dimensions}

We follow closely [4] and [9]. First, we note that the operators
$I_{P}, I_{T}$
and
$I_{PT}$
given by:
$$
I_{P}\cdot (x^{0},x^{1},x^{2}) = (x^{0},x^{1}, - x^{2})\eqno(3.1a)
$$
$$
I_{T}\cdot (x^{0},x^{1},x^{2}) = (- x^{0},x^{1},x^{2})\eqno(3.1b)
$$
$$
I_{PT}\cdot (x^{0},x^{1},x^{2}) = (- x^{0},x^{1}, - x^{2})\eqno(3.1c)
$$
are elements of the Lorentz group in 1+2 dimensions. They
belong respectively to
${\cal L}^{\uparrow}_{-}, {\cal L}^{\downarrow}_{-},
{\cal }L^{\downarrow}_{+}$.
The matrices
$I_{\tau}~~(\tau=0, P, T, PT)$
with
$I_{0} = {\bf 1}$
generates an Abelian discrete group  denoted by
$H_{inv}$.
Remark that unlike in 1+3 dimensions, the space inversion
$I_{S}\cdot (x^{0},x^{1},x^{2}) = (x^{0},- x^{1}, - x^{2})$
is an element of $\ol$
so it cannot be used as a representative for some connected component
of ${\cal L}$ different from $\ol$.

Next, we define for any
$h = (\phi,u) \in \ul$
and for any
$\tau = 0, P, T, PT$,
the element
$h_{\tau} \in \ul$
as follows:
$$
h_{0}=h,\eqno(3.2a)
$$
$$
(\phi,u)_{P}= (-\phi,\bar{u}),\eqno(3.2b)
$$
$$
(\phi,u)_{T} = (\phi,-u),\eqno(3.2c)
$$
$$
(\phi,u)_{PT}= (-\phi,-\bar{u})\eqno(3.2d)
$$
and prove the identity:
$$
I_{\tau} \delta(h) I_{\tau} = \delta(h_{\tau})\eqno(3.3)
$$
($\forall \tau =
0, P, T, PT,~ \forall h \in \ul$).

It is convenient to define
$\forall \tau = 0, P, T, PT$,
$s(\tau): H \rightarrow H$
by:
$$
s(\tau) h = h_{\tau}.\eqno(3.4)
$$

Then one easily shows that
$s(\tau)$
is an involutive automorphism of $H$ and moreover, the map
$\tau \mapsto s(\tau)$
is a group homomorphism:
$$
s(\tau)^{2} = Id\eqno(3.5a)
$$
$$
s(\tau) (h_{1}h_{2}) = s(\tau)(h_{1})\cdot s(\tau)(h_{2})\eqno(3.5b)
$$
$$
s(\tau_{1}\tau_{2}) = s(\tau_{1})\cdot s(\tau_{2})\eqno(3.5c)
$$

Now let $U$ be a projective isometric representation of ${\cal P}$. Then one
easily finds out that $U$ has the following structure:

- for
$\Lambda \in \ul$,
$U_{\Lambda,a} = {\cal W}_{h,a}$
where
$\delta(h) = \Lambda$
and ${\cal W}$ is a unitary representation of the inhomogeneous group
$in(\ul) \equiv \ul\times_{\delta} T(3)$
(see section 2D).

- for
$\Lambda = I_{\tau}$,
 $U_{I_{\tau},0} \equiv S_{\tau}$
verifies:
$$
S_{\tau}^{2} = c(\tau) \times {\bf 1},~~\vert c(\tau)\vert = 1\eqno(3.6)
$$
$$
S_{PT} = c S_{P} S_{T},~~\vert c\vert = 1.\eqno(3.7)
$$

Moreover, we must have:
$$
S_{\tau} {\cal W}_{h,a} S_{\tau}^{-1} = {\cal W}_{h_{\tau},I_{\tau}a}.
\eqno(3.8)
$$

Indeed, the proof from [4] ch. IX,6 is unchanged if we observe the fact that
$\op$ in 1+2 dimensions admits only one unitary one-dimensional representation,
namely the trivial representation (see [8]).

Now if we suppose like in [4] that
${\cal W}_{{\bf 1},a}$
is a non-trivial representation of
$T(3)$
and the corresponding projector valued measure has the support included in a
set of the form
$\{ p \in \R^{3} \vert p_{0} \geq c\}~~(c \in \R)$,
then it follows that
$S_{P}$
must be unitary and
$S_{T}, S_{PT}$
must be antiunitary. This means that the argument from [4] is unchanged and we
must have
$c(T),~c(PT) \in \{\pm 1\}$
and we can fix
$c = 1,~~c(P) = 1$
(see (3.6)-(3.8)). One usually calls
$(c(T),c(PT))$
the type of the representation $U$.

\subsection*{B Projective Unitary Representation of the Orthochronous
Poincar\'e Group}

{}From section A above it follows that if we restrict ourselves to the
orthochronous Poincar\'e group then we deal only with unitary operators.
We define now the group
$\tilde{H} \equiv {\bf Z}_{2}\times_{\tilde{s}} H~~(H = \ul)$
where:
$$
\tilde{s}_{1} = id,\eqno(3.9a)
$$
$$\tilde{s}_{-1}(h) = h_{P}\eqno(3.9b)
$$
and the semidirect product
$in(\tilde{H}) \equiv \tilde{H}\times_{\tilde{t}}T(3)$
where:
$$
\tilde{t}_{1,h}(a) = \delta(h)\cdot a\eqno(3.10a)
$$
$$
\tilde{t}_{-1,h}(a) = \delta(h) I_{P}\cdot a.\eqno(3.10b)
$$

Now, if
${\cal W}_{h,a}$
and
$S_{P}$
verify (3.6) + (3.8) (for
$\tau = P$
with
$c(P) = 1$)
we can define a unitary representation $W$ of
$in(\tilde{H})$
according to:
$$
W_{(1,h),a} = {\cal W}_{h,a},\eqno(3.11a)
$$
$$
W_{(-1,h),a} = {\cal W}_{h,a}S_{P}.\eqno(3.11b)
$$

Conversely, if $W$ is a unitary represaentation of
$\tilde{H}$
we can define
${\cal W}_{h,a}$
according to the first relation above and:
$$
S_{P} = W_{(-1,e_{H}),0};\eqno(3.12)
$$
in consequence, we will have (3.6) + (3.8). Moreover, it is clear that the
isometric representation $U$ of the orthochronous Poincar\'e group is
irreducible {\it iff} the unitary representation $W$ of
$in(\tilde{H})$
is irreducible.

Because
$in(\tilde{H})$
is a semidirect product between
$\tilde{H}$
and an Abelian group, we can apply the Mackey scheme 2A.

The orbits of
$\tilde{H}$
in
$\widehat{T(3)} \equiv \R^{3}$
are the same as in 2D and we choose the same representative points on them.

(I)
$$
\tilde{H}_{\eta e_{0}} = \{(\epsilon,(\phi,0))\vert \epsilon \in Z_{2},~\phi
\in \R\}
$$

Now we present in detail the method of obtaining the unitary irreducible
representations of
$\tilde{H}_{\eta e_{0}}$
because this case is generic. The origin of the anticipated "doubling"
phenomenon will be clear.

Let
$\tilde{D}(\epsilon,(\phi,0))$
be such a representation in ${\cal H}$. We define the unitary operator:
$$
J_{P} \equiv \tilde{D}(-1,e_{H})\eqno(3.13)
$$
and the unitary representation of the Abelian group $\R$:
$$
D(\phi) = \tilde{D}(1,(\phi,0)).\eqno(3.14)
$$

They are connected by:
$$
J_{P} D(\phi) J_{P}^{-1} = D(-\phi),\eqno(3.15)
$$
and we also have:
$$
J_{P}^{2} = Id.\eqno(3.16)
$$

According to SNAG theorem there exists a projector valued measure $P$ in ${\cal
H}$ based on
$\hat{\R} \simeq \R$
such that:
$$
D(\phi) = \int_{\R} e^{is\phi} dP(s).\eqno(3.17)
$$

Then (3.15) is equivalent to:
$$
P(-\Delta) = J_{P} P(\Delta) J_{P}^{-1}\eqno(3.18)
$$
for any Borel set
$\Delta \in \beta(\R)$.
Let us take now
$\Delta_{0} \in \beta(\R_{+}\cup \{0\})$
be fixed and let:
$$
{\cal H}_{0} \equiv \{f \in {\cal H}\vert [P(\Delta_{0}+P(-\Delta_{0})]f = f\}
\subset {\cal H}
$$

It is easy to see that
${\cal H}_{0}$
is invariant with respect to
$J_{P}$
and
$D(\phi)$,
so it is invariant with repect to the representation
$\tilde{D}$.
But
$\tilde{D}$
is supposed to be irreducible, so we have only two possibilities:

(i) there exists
$P_{0}$
a projector in ${\cal H}$
such that:
$$
P(\Delta) = \cases{P_{0}, & if $0 \in \Delta$\cr 0, & if $0
\not\in \Delta$.\cr}\eqno(3.19)
$$

It is clear that we have
${\cal H} = P_{0}{\cal H}$
so,
$$
D(\phi) = Id
$$

Then, from (3.16) it follows that
$J_{P}$
is an orthogonal projector. If we want
$\tilde{D}$
to be irreducible, the only way is to have
${\cal H} = {\bf C}$
and
$J_{P} = \lambda Id~~(\vert\lambda\vert = 1);$
(3.16) fixes now
$\lambda^{2} = 1$
i.e.
$\lambda = \pm 1$.
We obtain the irreducible representations
$\tilde{D}^{(\epsilon)}$
acting in ${\bf C}$ according to:
$$
\tilde{D}^{(+)}(\epsilon,(\phi,0)) = 1,\eqno(3.20a)
$$
$$
\tilde{D}^{(-)}(\epsilon,(\phi,0)) = \epsilon\eqno(3.20b)
$$

The corresponding induced representations are:
$\tilde{W}^{m,\eta,\epsilon} \equiv W^{\xm,D^{(\epsilon)}}$.
One notes that
$\tilde{W}^{m,\eta,+}$
and
$\tilde{W}^{m,\eta,-}$
induce the same representation of the orthochronous Poincar\'e group in the
logic
${\cal P}({\cal H})$,
so we can consider only the first one. One can take
${\cal H} = L^{2}(\xm,\mas)$
and we have (see (3.11) + (3.12)):
$$
\left(\tilde{\cal W}^{m,\eta,+}_{h,a} f\right)(p) = e^{i a\cdot p}
f(\delta(h)^{-1}\cdot p)\eqno(3.21a)
$$
$$
(S_{P} f)(p) = f(I_{P}\cdot p)\eqno(3.21b)
$$

(ii) there exists
$s \in \R_{+}$
and two orthogonal projectors
$P_{\pm}$
in ${\cal H}$ such that
$P_{+} P_{-} = 0$
and
$$
P(\Delta) = \cases{P_{\pm} & if $\pm s \in \Delta$, $\mp s \not\in \Delta$\cr
0 & if $s \not\in \Delta$, $-s \not\in \Delta$.\cr}\eqno(3.22)
$$

It is clear that
${\cal H} = {\cal H}_{+} \oplus {\cal H}_{-}$
where
${\cal H}_{\pm} \equiv P_{\mp} {\cal H}$
and we have in obvious matrix notations:
$$
D(\phi) = \left(\matrix{e^{-is\phi} {\bf 1} & 0\cr 0 & e^{is\phi}
{\bf 1}\cr}\right)
\eqno(3.23)
$$

{}From (3.18) we have
$J_{P} P_{\epsilon} = P_{-\epsilon} J_{P}$.
Combining with (3.16) we get:
$$
J_{P} = \left(\matrix{0 & I_{0}^{-1}\cr I_{0} & 0\cr}\right)
$$
with
$I_{0}: {\cal H}_{+} \rightarrow {\cal H}_{-}$
unitary. If we perform the unitary transformation
$U: {\cal H}_{+} \oplus {\cal H}_{-} \rightarrow {\cal H}_{+}
\oplus {\cal H}_{+}$
given by
$U(x,y) = (x, I_{0}^{-1}y)$,
then it follows that one can take
${\cal H} = {\cal H}_{+} \oplus {\cal H}_{+}$,
$D(\phi)$
remains in the same matrix form given by (3.23) and:
$$
J_{P} = \left(\matrix{0 & {\bf 1}\cr {\bf 1} & 0\cr}\right)\eqno(3.24)
$$

Finally, the irreducibility is compatible only with
${\cal H}_{+} = {\bf C}$.
The representation
$\tilde{D}^{(s)}$
acting in
${\bf C}^{2}$
is determined by (3.23) and (3.24). As a rule, we will denote
the vectors in
${\bf C}^{2}$
by
$v_{\pm}$
and similarly for functions with two components. The
corresponding induced representations
$\tilde{W}^{m,\eta,s} \equiv W^{\xm,\tilde{D}^{(s)}}$
can be realized according to the recipe (2.3) in
$L^{2}(\xm,\mas,{\bf C}^{2})$.
Using (3.11) + (3.12) we get:
$$
\left({\cal W}^{m,\eta,s}_{h,a} f\right)_{\pm}(p) =
e^{i a\cdot p} \varphi^{(\mp s)}(h,\delta(h)^{-1}\cdot p)
f_{\pm}(\delta(h)^{-1}\cdot p)\eqno(3.25a)
$$
$$
(S_{P} f)_{\pm}(p) = f_{\mp}(I_{P}\cdot p)\eqno(3.25b)
$$

It is important that
${\cal W}^{m,+,s}$
can also be realized in the fiber bundle formalism as in 2B, 2D.
One uses the Hilbert space
$\tilde{\cal H}^{+,s}_{m} \equiv {\cal H}^{+,-s}_{m} \oplus
{\cal H}^{+,s}_{m} $
and has:
$$
\left({\cal W}^{m,+,s}_{h,a} f\right)_{\pm}(p) =
e^{i a\cdot p} D^{(s/2,\pm)}(h)\cdot f_{\pm}(\delta(h)^{-1}\cdot p)\eqno(3.26a)
$$
$$
(S_{P} f)_{\pm}(p) = f_{\mp}(I_{P}\cdot p)\eqno(3.26b)
$$

The proof goes as in section 2D. One defines the following action
$\tilde{D}$
of
$\tilde{H}$
in the fiber bundle space
$\tilde{B}^{+,s}_{m} \equiv {B}^{+,-s}_{m} \oplus {B}^{+,s}_{m}$:
$$
\tilde{D}(1,h)\cdot (p,f_{\pm}) = (\delta(h)\cdot p,
D^{(s/2,\pm)}(h)\cdot f_{\pm})\eqno(3.27a)
$$
$$
\tilde{D}(-1,e_{H})\cdot (p,f_{\pm}) = (I_{P}\cdot p,f_{\mp})\eqno(3.27b)
$$
(compare with (2.26)). To check that
$\tilde{D}$
is well defined is easy: for elements of the type
$(1,h)$
one uses, as before, (2.22) and for
$(-1,e_{H})$
one has to convince oneself that
$H^{(-\epsilon)}_{0} = - H^{(\epsilon)}_{0},~~
H^{(-\epsilon)}_{1} = - H^{(\epsilon)}_{1}$
and
$H^{(-\epsilon)}_{2} = H^{(\epsilon)}_{2}$.
The group property of
$\tilde{D}$
follows from (3.3) and the identity
$$
D^{(l,\epsilon)}(h) = D^{(l,-\epsilon)}(h_{P})\eqno(3.28)
$$
(see (2.12) and (2.13)). Finally, the fiber
$\tilde{B}^{+,s}_{m}(me_{0})$
is two dimensional and we have explicitely:
$\tilde{B}^{+,s}_{m}(me_{0}) = (me_{0},c_{\pm})$
where
$c_{\pm}$
are constant functions in $F$. It is clear from (3.27) that:
$$
\tilde{D}(1,(\phi,0))\cdot (me_{0},c_{\pm}) = (me_{0},
e^{\mp is\phi} c_{\pm})
$$
$$
\tilde{D}(-1,e_{H})\cdot (me_{0},c_{\pm}) = (me_{0},c_{\mp})
$$
i.e. we have obtained the representation
$\tilde{D}^{(s)}$
determined by (3.23) and (3.24).

(II)

$$
\tilde{H}_{\eta e_{+}} = \left\{ \left(\epsilon,\left( {1 \over 2i}
ln {1-ib \over 1+ib} +n\pi, {ib \over 1-ib} \right)\right) \vert
\epsilon \in Z_{2}, n \in Z, b \in \R\right\}.
$$

Let
$\tilde{D}$
be a unitary irreducible representation of
$\tilde{H}_{\eta e_{+}}$.
We define the unitary operator
$$
J_{P} = \tilde{D}(-1,e_{H})\eqno(3.29)
$$
and the unitary representation of the Abelian group
${\bf Z} \times \R$:
$$
D(n,b) = \tilde{D} \left(1,\left( {1 \over 2i} ln {1-ib \over 1+ib}
+n\pi, {ib \over 1-ib} \right)\right).\eqno(3.30)
$$

They are connected by:
$$
J_{P} D(n,b) J_{P}^{-1} = D(-n,-b)\eqno(3.31)
$$
and we also have:
$$
J_{P}^{2} = Id.\eqno(3.32)
$$

Again SNAG theorem gives us a projector valude measure $P$ in
${\cal H}$
based on
$\widehat{{\bf Z} \times \R} \simeq T^{1} \times \R$
($T^{1}$
is the torus
$\vert z\vert = 1$
in ${\bf C}$) such that:
$$
D(n,b) = \int z^{n} e^{isb} dP(z,s).\eqno(3.33)
$$

{}From (3.31) we have equivalently:
$$
J_{P} P(\Delta_{1},\Delta_{2}) J_{P}^{-1} = P(\overline{\Delta_{1}},
-\Delta_{2}),\eqno(3.34)
$$
($\forall \Delta_{1} \in \beta(T^{1}),~\forall
\Delta_{2} \in \beta(\R)$.

The same irreducibility argument from case (I) leads us to three
cases:

(i) there exists an orthogonal projector
$P_{0}$
in ${\cal H}$ such that:
$$
P(\Delta) = \cases{ P_{0}, & if $(1,0) \in \Delta$\cr 0 & if
$(1,0) \not\in \Delta$.\cr}\eqno(3.35)
$$

In this case
$P_{0} {\cal H} = {\cal H}$
and one has:
$$
D(n,b) = Id.\eqno(3.36)
$$

As in case (I) above one finds out that in fact
${\cal H} = {\bf C}$
and besides (3.36)
we also have:
$$
J_{P} = \epsilon\times Id~~(\epsilon = \pm).\eqno(3.37)
$$

We denote the corresponding representations
$\tilde{D}^{(+,\epsilon)}$.

(ii) there exists an orthogonal projector
$P_{0}$
in ${\cal H}$ such that:
$$
P(\Delta) = \cases{ P_{0}, & if $(-1,0) \in \Delta$\cr 0 & if
$(-1,0) \not\in \Delta$.\cr}\eqno(3.38)
$$

Then, as before, one obtains
${\cal H} = {\bf C}$,
$$
D(n,b) = (-1)^{n}\times Id\eqno(3.39a)
$$
$$
J_{P} = \epsilon \times Id\eqno(3.39b)
$$

We denote these representations by
$\tilde{D}^{(-,\epsilon)}$.

(iii) there exist
$(z,s) \in T^{1}\times (\R_{+}\cup\{0\}) -\{\pm 1,0\}$
and
$P_{\pm}$
two orthogonal projectors in ${\cal H}$ such that
$P_{+}P_{-} = 0$
and:
$$
P(\Delta) = \cases{ P_{+} &  if $(z,s) \in \Delta,~(\bar{z},-s)
\not\in \Delta$,\cr P_{-} & if $(\bar{z},-s) \in \Delta,~(z,s)
\not\in \Delta$,\cr 0 &  if $(z,s) \not\in \Delta,~(\bar{z},-s)
\not\in \Delta$\cr}\eqno(3.40)
$$

One finds out that
${\cal H} = {\bf C} \oplus {\bf C}$
and:
$$
D(n,b) = \left(\matrix{\bar{z}^{n} e^{-isb} {\bf 1} & 0\cr 0 &
z^{n} e^{isb} {\bf 1}\cr}\right)\eqno(3.41a)
$$
$$
J_{P} = \left(\matrix{0 & {\bf 1}\cr {\bf 1} & 0\cr}\right)\eqno(3.41b)
$$

We denote this representation by
$\tilde{D}^{(z,s)}$.
The corresponding induced representations are denoted:
$W^{\eta,\epsilon,\epsilon'} \equiv
W^{\xf,\tilde{D}^{(\epsilon,\epsilon')}}$
and
$W^{\eta,s,t} \equiv W^{\xf,\tilde{D}^{(e^{i\pi s},t)}}$;~
$W^{\eta,\epsilon,+}$
and
$W^{\eta,\epsilon,-}$
induce the same representations in the logic
${\cal P}({\cal H})$.
For the explicit expressions we have two options: to use the
canonical  formalism from 2A or to use the fiber bundle
formalism. In the first case one has to use in an obvious way
the cocycles
$\varphi^{(\pm s,\pm t)}$
(see 2D(II)). The fiber bundle trick seems to work only if
$s \not= 0,1$
and
$t = 0$.
The formul\ae~are formally identical to (3.26).

(III)
$$
\tilde{H}_{me_{2}} = \left\{\left(1,\left( \pi n,{a-1\over a+1}\right)
\right)\vert n \in {\bf Z}, a \in \R_{+} \right\}\cup
$$
$$
\left\{\left(-1,\left({\pi\over 2}+\pi n,-{a-1\over a+1}\right)
\right)\vert n \in {\bf Z}, a \in \R_{+} \right\}.
$$

We do not insist on this case because we will, not need it
later. We only mention that one can use many facts from case
(II) if one convinces oneself that the map
$\pi:\tilde{H}_{me_{2}} \rightarrow \tilde{H}_{\eta e_{+}}$
given by:
$$
\pi\left(1,\left( \pi n,{a-1\over a+1}\right)
\right) = \left(1,\left( {1 \over 2i} ln {1-iln(a) \over 1+iln(a)}
+n\pi, {iln(a) \over 1-iln(a)} \right)\right)
$$
$$
\pi\left(-1,\left({\pi\over 2}+\pi n,-{a-1\over a+1}\right)
\right) = \left(-1,\left( {1 \over 2i} ln {1-iln(a) \over 1+iln(a)}
+n\pi, {iln(a) \over 1-iln(a)} \right)\right)
$$
is a group homomorphism. This observation easily provides the
unitary irreducible representations of
$\tilde{H}_{me_{2}}$.
The induced representations follow as above.

{\bf Remark 2:} A byproduct of the analysis above is the fact
that the projective unitary irreducible representations of $\op$
corresponding to
$s \not= 0$
do not admit extensions to the orthochronous Poincar\'e group.

\subsection*{C Mixed Representations}

The case of the full Poincar\'e group ${\cal P}$ can be analysed
with the help of the notions of mixed representation, mixed
systems of imprimitivity, etc. [9]. The idea is to replace
everywhere the group
$M({\cal H})$
of unitary operators in the Hilbert space ${\cal H}$ with the group
$\bar{M}({\cal H})$
of unitary and antiunitary operators in ${\cal H}$. For
instance, a mixed representation of the group $G$ will be a
continous homomorphism
$U: G \rightarrow \bar{M}({\cal H})$.

Let
$H\times_{t} A$
be a semidirect product of $H$ and $A$ as in 2A. We are
interested in the classification of the mixed irreducible
representations of
$H\times_{t} A$
verifying
$$
U_{e_{H},a} = unitary,~~\forall a \in A.\eqno(3.42)
$$

Let us denote
$H^{u} \equiv \{h \in H\vert U_{h,a} = unitary\}$
and
$H^{a} \equiv \{h \in H\vert U_{h,a} = antiunitary\}$.

One can show that the Mackey procedure from 2A works in this
case also with some technical changes.

(a) As in 2A,
$\hat{A}$
is the dual of $A$. For any
$\omega \in \hat{A}$
we define
$\omega^{*} \in \hat{A}$
by:
$$
\omega^{*}(a) = \overline{\omega(a)}.\eqno(3.43)
$$

Next, we have an action of $H$ in
$\hat{A}$
given by
$$
(h,\omega) \rightarrow \cases{h\cdot \omega & if
$h \in H^{u}$ \cr h\cdot \omega^{*} & if $h \in H^{a}$\cr}.\eqno(3.44)
$$
(For the definition of
$h\cdot\omega$
see (2.1)).

(b), (c) are the same steps as in 2A i.e. one computes the
$H$-orbits relative to the action above, choose some reference
points
$\omega_{0} \in Z$
and computes the "little" groups
$H_{\omega_{0}}$.

(d) One tries to find out a complete list of all mixed
representations
$\pi$
of
$H_{\omega}$
verifying the condition:
$$
\pi (h) = \cases{unitary & $h \in H^{u}$\cr
antiunitary & $h \in H^{a}$\cr}.\eqno(3.45)
$$

(e) For every such mixed representation
$\pi$
acting in ${\cal H}$
and corresponding to the orbit
${\cal O}$
one tries to find an
$(H,{\cal O},\bar{M}({\cal H}))$-
cocycle
$\varphi^{\pi}$
which is admissible i.e.
$\varphi^{\pi}(h,\omega) \varphi^{\pi}(h,\omega')$
is always unitary.

(f) The induction procedure 2A(f) is unchanged and in particular
we have (2.3).

\subsection*{D Projective Mixed Representations}

First, we have to exhibit somehow a semidirect product
structure. To this purpose it is convenient to rewrite
(3.6)-(3.8) (with
$c = c(P) = 1,~~c(T) = \epsilon_{T}~~c(PT) =
\epsilon_{PT},~~\epsilon_{T}, \epsilon_{PT} \in \{\pm 1\})$
as follows:

$$
{\cal W}_{h,a} S_{\tau} {\cal W}_{h',a'} S_{\tau'} =
c(\tau,\tau')~{\cal W}_{(h,a)\cdot(h_{\tau}',I_{\tau}a')}~
S_{\tau\circ\tau'}\eqno(3.46)
$$
where
$c(\cdot,\cdot)$
is a multiplicator for the group
$H_{inv}$
given by:
$$
c(\tau,\tau') = \cases{\epsilon_{T} & if $\tau = \tau' = T$, or
$\tau = PT, \tau' =T$\cr \epsilon_{PT} & if $\tau = \tau' = PT$, or
$\tau = T, \tau' =PT$\cr \epsilon_{T}\epsilon_{PT} & if $\tau = T,
\tau' = P$, or $\tau = PT, \tau' =P$\cr 0 & in other cases \cr}.\eqno(3.47)
$$

Next, we define the group
$\tilde{\tilde{H}} = (Z_{2} \times Z_{2})\times_{\tilde{\tilde{s}}} H$
where:
$$
\tilde{\tilde{s}}_{1,1} = id,~~\tilde{\tilde{s}}_{1,-1}(h) = h_{P},~~
\tilde{\tilde{s}}_{-1,1}(h) = h_{T},~~\tilde{\tilde{s}}_{-1,-1}(h) = h_{PT}.
\eqno(3.48)
$$
and the semidirect product
$in(\tilde{\tilde{H}}) \equiv \tilde{\tilde{H}}\times_{\tilde{\tilde{t}}} T(3)$
where:
$$
\tilde{\tilde{t}}_{(1,1),h}(a) = \delta(h) a\eqno(3.49a)
$$
$$
\tilde{\tilde{t}}_{(1,-1),h}(a) = \delta(h)I_{P}\cdot a\eqno(3.49b)
$$
$$
\tilde{\tilde{t}}_{(-1,1),h}(a) = \delta(h)I_{T}\cdot a\eqno(3.49c)
$$
$$
\tilde{\tilde{t}}_{(-1,-1),h}(a) = \delta(h)I_{PT}\cdot a.\eqno(3.49d)
$$

Finally we define:
$$
W_{((1,1),h),a} = {\cal W}_{h,a}\eqno(3.50a)
$$
$$
W_{((1,-1),h),a} = {\cal W}_{h,a} S_{P}\eqno(3.50b)
$$
$$
W_{((-1,1),h),a} = {\cal W}_{h,a} S_{T}\eqno(3.50c)
$$
$$
W_{((-1,-1),h),a} = {\cal W}_{h,a} S_{PT}\eqno(3.50d)
$$

Then one can easily show that $W$ is a projective mixed
representation of
$\tilde{\tilde{H}}$
with the multiplicator
$\tilde{\tilde{c}}$
given by:
$$
\tilde{\tilde{c}}((\tilde{\tilde{h}}_{1},a_{1}),
(\tilde{\tilde{h}}_{2},a_{2})) =
c(p(\tilde{\tilde{h}}_{1}),p(\tilde{\tilde{h}}_{2}))
\equiv \tilde{\tilde{c}}(\tilde{\tilde{h}}_{1},\tilde{\tilde{h}}_{2})
\eqno(3.51)
$$
with
$$
p((1,1),h) = 0,~~ p((1,-1),h) = P,~~ p((-1,1),h) = T,~~p((-1,-1),h) =
PT.
$$

In this case
$\tilde{\tilde{H}}^{u} \equiv \{ ((1,\epsilon),h)\vert \epsilon \in Z_{2},~ h
\in H\}$
and
$\tilde{\tilde{H}}^{a} \equiv \{ ((-1,\epsilon),h)\vert \epsilon \in Z_{2},~ h
\in H\}$.

Conversely if $W$ is such a projective mixed representation of
$\tilde{\tilde{H}}$,
then the operators:
$$
{\cal W}_{h,a} \equiv W_{((1,1),h),a}\eqno(3.52a)
$$
$$
S_{P} \equiv W_{((1,-1),e_{H}),0}\eqno(3.52b)
$$
$$
S_{T} \equiv W_{((-1,1),e_{H}),0}\eqno(3.52c)
$$
$$
S_{PT} \equiv W_{((-1,-1),e_{H}),0}\eqno(3.52d)
$$
verify (3.46) with
${\cal W}_{h,a}$
and
$S_{P}$
unitary and
$S_{T},~S_{PT}$
antiunitary. Moreover, the isometric representation $U$ of
${\cal P}$ we started with (see section A) is irreducible {\it
iff} $W$ above is irreducible.

We still do not have a true (mixed) representation of a
semidirect product so we still cannot apply Mackey procedure.
One has a standard trick to obtain from a projective
representation a true representation of the centrally extended
group. Namely, one defines
$\tilde{\tilde{H}}^{c} \equiv \{(\tilde{\tilde{h}},\epsilon)\vert
\tilde{\tilde{h}} \in \tilde{\tilde{H}}, \epsilon \in Z_{2}\}$
with the composition law:
$$
(\tilde{\tilde{h}}_{1},\epsilon_{1})\cdot
(\tilde{\tilde{h}}_{2},\epsilon_{2}) =
(\tilde{\tilde{h}}_{1}\tilde{\tilde{h}}_{2},
\tilde{\tilde{c}}(\tilde{\tilde{h}}_{1},\tilde{\tilde{h}}_{2})
\epsilon_{1}\epsilon_{2})\eqno(3.53)
$$
and the semidirect product
$in(\tilde{\tilde{H}}^{c}) \equiv \tilde{\tilde{H}}^{c}
\times_{\tilde{\tilde{t}}^{c}} T(3)$
where:
$\tilde{\tilde{t}}^{c}(\tilde{\tilde{h}},\epsilon) =
\tilde{\tilde{t}}_{\tilde{\tilde{h}}}$
(see (3.49)).

Let now $W$ be a projective mixed representation of
$\tilde{\tilde{H}}$
with the multiplier
$\tilde{\tilde{c}}$.
We define:
$$
W^{c}_{(\tilde{\tilde{h}},\epsilon),a} \equiv \epsilon
W_{\tilde{\tilde{h}},a}\eqno(3.54)
$$
and one can verify that
$W^{c}$
is a true mixed representation of the group
$in(\tilde{\tilde{H}}^{c})$
verifying:
$$
W^{c}_{(((1,1),e_{H}),\epsilon''),0} = \epsilon'' {\bf 1}\eqno(3.55)
$$
and
$$
W^{c}_{(((\epsilon,\epsilon'),h),\epsilon''),a} =
\cases{unitary & $\epsilon'' = +$\cr
antiunitary & $\epsilon'' = -$\cr}.\eqno(3.56)
$$

Conversely, if
$W^{c}$
is such a representation, then
$$
W_{\tilde{\tilde{h}},a} \equiv W^{c}_{(\tilde{\tilde{h}},1),a}\eqno(3.57)
$$
is a projective mixed representation with
$H^{u}$
and
$H^{a}$
as above and with the multiplier
$\tilde{\tilde{c}}$.

Moreover, the isometric representation $U$ of ${\cal P}$ we have
started with is irreducible {\it iff}
$W^{c}$
is irreducible.

Finally, we have succeeded to reduce the analysis to the
application of the scheme from section C. According to (3.45) in
step (d) one must look for irreducible representations of the
"little" groups such that:
$$
\pi_{((\epsilon,\epsilon'),h),\epsilon''} =
\cases{unitary & $\epsilon'' = +$\cr
antiunitary & $\epsilon'' = -$\cr}.\eqno(3.58)
$$

Also, one can see that the condition (3.55) will be fulfilled
{\it iff}:
$$
\pi_{((1,1),e_{H}),\epsilon''} = \epsilon'' {\bf 1}.\eqno(3.59)
$$

To summarize, one completes step (d) taking care of (3.58) +
(3.59), applies the induction procedure (f) and obtains the
desired projective isometric representation $U$ of ${\cal P}$ as
follows:
$$
U_{\Lambda,a} = W^{c}_{(((1,1),h),1),a}~~with~\delta(h) = \Lambda\eqno(3.60a)
$$
$$
S_{P} = W^{c}_{(((1,-1),e_{H}),1),0}\eqno(3.60b)
$$
$$
S_{T} = W^{c}_{(((-1,1),e_{H}),1),0}\eqno(3.60c)
$$
$$
S_{PT} = W^{c}_{(((-1,-1),e_{H}),1),0}.\eqno(3.60d)
$$

Because of (3.7) one needs to display only the first three lines.

\subsection*{E Projective Mixed Irreducible Representations of
${\cal P}$}

It is appearent from section D above that we have to apply the
modified Mackey procedure from section C to the group
$\tilde{\tilde{H}}^{c}$.

In the usual identification
$\widehat{T(3)} \equiv \R^{3}$,
the action (3.44) is:
$$
(((1,1),h),\epsilon'')\cdot [p] = [\delta(h)\cdot p]\eqno(3.61a)
$$
$$
(((1,-1),h),\epsilon'')\cdot [p] = [\delta(h)I_{P}\cdot p]\eqno(3.61b)
$$
$$
(((-1,1),h),\epsilon'')\cdot [p] = [\delta(h)I_{S}\cdot p]\eqno(3.61c)
$$
$$
(((-1,-1),h),\epsilon'')\cdot [p] = [\delta(h)I_{1}\cdot p].\eqno(3.61d)
$$

Here
$$
I_{S}\cdot (x^{0},x^{1},x^{2}) = (x^{0},-x^{1},-x^{2})
$$
is the spatial inversion and
$$
I_{1}\cdot (x^{0},x^{1},x^{2}) = (x^{0},-x^{1},x^{2}).
$$

It follows that the orbits remain the same and we choose the
same representative points on them. We sketch briefly the
implementation of the steps (c)-(f) of the Mackey procedure.

(I)
$(H^{c})_{\eta me_{0}} = \{(((\epsilon,\epsilon'),
(\phi,0)),\epsilon'') \vert \phi \in \R,~\epsilon, \epsilon',\epsilon''
\in Z_{2}\}$

Let
$\tilde{\tilde{D}}$
be a mixed irreducible representation of this subgroup verifying
(3.58) and (3.59). We define as in section B:
$$
J_{P} \equiv \tilde{\tilde{D}}(((1,-1),e_{H}),1)\eqno(3.62a)
$$
$$
J_{T} \equiv \tilde{\tilde{D}}(((-1,1),e_{H}),1)\eqno(3.62b)
$$
$$
J_{PT} \equiv \tilde{\tilde{D}}(((-1,-1),e_{H}),1)\eqno(3.62c)
$$
and
$$
D(\phi) \equiv \tilde{\tilde{D}}(((1,1),(\phi,0)),1).\eqno(3.63)
$$

Note that
$J_{P},~D(\phi)$
are unitary and
$J_{T}, J_{PT}$
are antiunitary. We still have (3.15) and (3.16) and also:
$$
J_{T} D(\phi) J_{T}^{-1} = D(\phi)\eqno(3.64)
$$
$$
J_{T}^{2} = \epsilon_{T} {\bf 1}\eqno(3.65a)
$$
$$
J_{PT}^{2} = \epsilon_{PT} {\bf 1}\eqno(3.65b)
$$
$$
J_{P} J_{T}  = \epsilon_{T}\epsilon_{PT} J_{T} J_{P}.\eqno(3.65c)
$$

It is not hard to see that we again have the same possibilities
as in section B (I):

(i) $P(\Delta)$
is given by (3.19). We must have
$\epsilon_{T} = \epsilon_{PT} = 1$
and we get
${\cal H} = {\bf C}$
with
$D(\phi) = {\bf 1},~J_{P} = \epsilon {\bf 1},~J_{T} = \pm K $,
($K$ is the complex conjugation). Applying the unitary transformation
$U = i {\bf 1}$
we can make
$J_{T} = K$.
The corresponding representations are denoted by
$\tilde{\tilde{D}}^{(\epsilon)}$
and the corresponding induced representations by
$\tilde{\tilde{W}}^{m,\eta,\epsilon}$;~
it is sufficient to consider
$\tilde{\tilde{W}}^{m,\eta,+}$.
Using (3.52) + (3.57) we have the same Hilbert space
$L^{2}(\xmp,\masp)$
and (3.21) must be supplemented by:
$$
(S_{T}f)(p) = \overline{f(I_{S}p)}.\eqno(3.66)
$$

(ii) $P(\Delta)$
is given by (3.22). One finds out that it is necessary to have
$\epsilon_{PT} = 1$
and we get
${\cal H} = {\bf C}$
with
$D(\phi)$
and
$J_{P}$
given by (3.23) + (3.24); also:
$$
J_{T} = \left(\matrix{0 & 1\cr \epsilon_{T} & 0\cr}\right) K.\eqno(3.67)
$$

The corresponding representations are denoted by
$\tilde{\tilde{D}}^{(s)}$
and the corresponding induced representations by
$\tilde{\tilde{W}}^{m,\eta,s}$.
These representations can be realized without changing the
Hilbert space from section B. One must add to (3.25);
$$
(S_{T} f)_{+} = \overline{f_{-}(I_{S}\cdot p)}\eqno(3.68a)
$$
$$
(S_{PT } f)_{-} = \epsilon_{T} \overline{f_{+}(I_{S}\cdot p)}.\eqno(3.68b)
$$

Alternatively, in the fiber bundle formalism, we can consider
the case
$\epsilon_{T} = 1$; for this one must add to (3.26):
$$
(S_{T} f)_{\pm} (p) = IC f_{\mp}(I_{S}\cdot p).\eqno(3.69)
$$

Here, the operation
$C: F \rightarrow F$
has been defined at (2.14) and
$I: F \rightarrow F$
is:
$$
(I f)(z) = f(-z).\eqno(3.70)
$$

(II)
$$
\tilde{\tilde{H}}_{\eta e_{+}} =
\left\{ \left(\left((1,\epsilon'),\left( {1 \over 2i}
ln {1-ib \over 1+ib} +n\pi, {ib \over 1-ib}
\right)\right),\epsilon''\right)  \vert
\epsilon', \epsilon'' \in Z_{2}, n \in Z, b \in \R\right\}\cup
$$
$$
\left\{ \left(\left((-1,\epsilon'),\left( {1 \over 2i}
ln {1-ib \over 1+ib} +{\pi\over 2}+n\pi,-{ib \over 1-ib}
\right)\right),\epsilon''\right)  \vert
\epsilon', \epsilon'' \in Z_{2}, n \in Z, b \in \R\right\}.
$$

We define as before:
$$
J_{P} \equiv \tilde{\tilde{D}}(((1,-1),e_{H}),1)\eqno(3.71a)
$$
$$
J_{T} \equiv \tilde{\tilde{D}}(((-1,1),e_{H}),1)\eqno(3.71b)
$$
$$
J_{PT} \equiv \tilde{\tilde{D}}(((-1,-1),e_{H}),1)\eqno(3.71c)
$$
and
$$
D(n,b) \equiv \tilde{\tilde{D}}\left(\left((1,1),\left( {1 \over 2i}
ln {1-ib \over 1+ib} +n\pi, {ib \over 1-ib} \right)\right),1\right)
\eqno(3.72)
$$

We have beside (3.31) and (3.32):
$$
J_{T} D(n,b) J_{T}^{-1} = D(n,b),\eqno(3.73)
$$
$$
J_{T}^{2} = \epsilon_{T} D(1,0)\eqno(3.74a)
$$
$$
J_{PT}^{2} = \epsilon_{PT} {\bf 1}\eqno(3.74b)
$$
$$
J_{P} J_{T}  = \epsilon_{T}\epsilon_{PT} D(-1,0) J_{T} J_{P}\eqno(3.74c)
$$

The irreducibility argument leads us again to the three cases
from B (II).

(i) $P(\Delta)$
is given by (3.35). We must have
$\epsilon_{T} = \epsilon_{PT} = 1$.
We get
${\cal H} = {\bf C},~D(n,b) = {\bf 1},~J_{P} = \epsilon {\bf 1}$
and
$J_{T} = K$.
The corresponding representation are denoted by
$\tilde{\tilde{D}}^{(+,\epsilon)}$.

(ii) $P(\Delta)$
is given by (3.38). One must have
$\epsilon_{T} = - \epsilon_{PT} = - 1$.
We get
${\cal H} = {\bf C},~D(n,b) = (-1)^{n} {\bf 1},~J_{P} = \epsilon {\bf 1}$
and
$J_{T} = K$.
We denote these representations by
$\tilde{\tilde{D}}^{(-,\epsilon)}$.

(iii) $P(\Delta)$
is given by (3.40). We must have
$\epsilon_{PT} = 1$.
Then
${\cal H} = {\bf C}^{2}$
and beside (3.41) one has:
$$
J_{T} = \left(\matrix{0 & 1\cr \epsilon_{T} z & 0\cr}\right) K.\eqno(3.75)
$$
These representations are denoted by
$\tilde{\tilde{D}}^{(s)}$.
The induced representations
$\tilde{\tilde{W}}^{\eta,\epsilon,\epsilon'},
\tilde{\tilde{W}}^{\eta,s,t}$
can be easily realized in the cannonical formalism (2.3).

Especially interesting is the case (iii) for
$\epsilon_{T} = 1$,
$t = 0$
and
$s \not= 0,1$
when we can use the fiber bundle formalism. Indeed, one simply makes
$m \rightarrow 0$
in the formul\ae~(3.26) + (3.69).

\section{Free Relativistic Fields}

\subsection*{A The Fock-Cook Formalism [15], [16]}

The formalism of the second quantization, due mainly to Fock and
Cook, is a natural way to associate to a single particle system
a multi-particle system. We present the main ingredients which
will be needed in the following.

Let
$({\cal H},(\cdot,\cdot))$
be a separable complex Hilbert space. Then we define the
associated Fock space:
$$
\foc \equiv \oplus_{n=0}^{\infty} {\cal H}^{(n)}\eqno(4.1)
$$
where
${\cal H}^{(0)} = {\bf C},~{\cal H}^{(n)} = {\cal H}^{\otimes n}$.
The algebraic Fock space is:
$$
\foc^{0} \equiv \{ \Phi^{(n)} \in \foc\vert~\exists N \in \N
{}~s.t.~\Phi^{(n)} = 0,~\forall n>N\};
$$
$\foc^{0}$
is dense in $\foc$.

We define on the algebraic Fock space the annihilation and the
creation operators: for any
$f \in {\cal H}$
these operators are denoted by
$b(f)$
and
$b^{*}(f)$
respectively and they act on decomposable vectors according to:
$$
b(f)~\otimes_{i=1}^{n} f_{i} = \sqrt{n}~(f,f_{1})~\otimes_{i=2}^{n}
f_{i}\eqno(4.2)
$$
$$
b^{*}(f)~\otimes_{i=1}^{n} f_{i} = \sqrt{n+1}~f~\otimes_{i=1}^{n}
f_{i}.\eqno(4.3)
$$

These relations are valid for any $n$ if one uses the Bourbaki convention:
$\otimes_{i \in \Phi} = {\bf 1}$.
One extends
$b(f)$
and
$b^{*}(f)$
to
$\foc^{0}$
by
linearity and continuity.
One also notices that:
$$
(\Psi,b^{*}(f) \Phi) = (b(f) \Psi,\Phi).\eqno(4.4)
$$

Next, one defines the symmetrization and the anti-symmetrization
projectors:
$$
P_{+} \equiv \oplus_{n=0}^{\infty} S_{n},~~
P_{-} \equiv \oplus_{n=0}^{\infty} A_{n}\eqno(4.5)
$$
where:
$$
S_{n} \otimes_{i=1}^{n} f_{i} = \sum_{\sigma \in P_{n}}
\otimes_{i=1}^{n} f_{\sigma^{-1}(i)}\eqno(4.6)
$$
$$
A_{n} \otimes_{i=1}^{n} f_{i}= \sum_{\sigma \in P_{n}}
(-1)^{\vert\sigma\vert} \otimes_{i=1}^{n} f_{\sigma^{-1}(i)}.\eqno(4.7)
$$

Here
$P_{n}$
is the permutation group of
$\{1,2,\cdots,n\}$
and
$\vert\sigma\vert$
is the signature of the permutation
$\sigma$.

Then
${\cal H}_{\pm} \equiv P_{\pm} {\cal H}$
are called the Bosonic and respectively the Fermionic Fock
spaces. The operators:
$$
a_{\pm}(f) \equiv P_{\pm} b(f) P_{\pm}\eqno(4.8a)
$$
$$
a^{*}_{\pm}(f) \equiv P_{\pm} b^{*}(f) P_{\pm}\eqno(4.8b)
$$
are called the Bosonic (Fermionic) annihilation and creation
operators. One can prove the following formul\ae:
$$
[a_{\pm}(f),a_{\pm}(g)]_{\pm} = 0\eqno(4.9a)
$$
$$
[a^{*}_{\pm}(f),a^{*}_{\pm}(g)]_{\pm} = 0\eqno(4.9b)
$$
$$
[a_{\pm}(f),a^{*}_{\pm}(g)]_{\pm} = (f,g) {\bf 1}_{\foc}\eqno(4.9c)
$$
(the (anti)-cannonical commutation relations).

We can formulate now the correspondence between some
one-particle system and the associated multi-particle system. If
the one-particle system is described by the lattice
${\cal P}({\cal H})$,
then to every element
$Q \in {\cal P}({\cal H})$
($Q$ is any orthogonal projector)
one associates
$d\Gamma(Q) \in {\cal P}({\cal F}_{\pm}({\cal H}))$
according to:
$$
d\Gamma(Q) \equiv Q \otimes {\bf 1}_{\cal H} \otimes \cdots
\otimes {\bf 1}_{\cal H} + {\bf 1}_{\cal H} \otimes Q \otimes\cdots
\otimes {\bf 1}_{\cal H} + \cdots + {\bf 1}_{\cal H} \otimes
\cdots \otimes {\bf 1}_{\cal H} \otimes Q.\eqno(4.10)
$$

With some domain precautions, this formula also works for
"observables", i.e. arbnitrary self-adjoint operators in ${\cal H}$.

For any
$f \in {\cal H}$
one can define the Segal field operator
$\Phi^{\pm}(f)$
on
$\foc^{0}$
by:
$$
\Phi^{\pm}(f) = {1\over\sqrt{2}} [a_{\pm}(f) + a^{*}_{\pm}(f)].\eqno(4.11)
$$

The motivation of this definition is the usual connection
appearing for the harmonic oscillator
between, on the one hand the creation and annihilation operators
and, on the other hand the position operator.

Let us suppose that
$U: {\cal H} \rightarrow {\cal H}$
is a unitary or anti-unitary operator. Then one defines
$\Gamma(U): \foc \rightarrow \foc$
by:
$$
\Gamma(U) \equiv \sum_{n=0}^{\infty} U^{\otimes n}.\eqno(4.12)
$$

It is clear that
$\Gamma(U)$
is (anti)-unitary {\it iff} $U$ is (anti)-unitary. Suppose we
have the time evolution of the one-particle system in ${\cal H}$
i.e. we have a unitary representation
$t \mapsto U_{t}$
of the Abelian group $\R$ in ${\cal H}$. Then the evolution of
the multi-particle system is dictated by the unitary
representation
${\cal U}_{t} \equiv \Gamma(U_{t})$.
Because
$$
\Gamma(U_{1} U_{2}) = \Gamma(U_{1}) \Gamma(U_{2})\eqno(4.13)
$$
this definition is consistent. In this case one can define the
Segal field at time $t$ as follows:
$$
\Phi^{\pm}_{t}(f) \equiv {\cal U}_{t} \Phi^{\pm}(f)
{\cal U}_{t}^{-1}.\eqno(4.14)
$$

Suppose now that we are given the one-particle Hilbert space
${\cal H}$ and the statistics: Bose or Fermi. Can we construct
in a natural way some Wightman quantum free field living in
${\cal F}_{\pm}({\cal H})$?
{}From (4.14) it is clear that one needs
$\Phi({\bf x})$
and
$U_{t}$.
If in ${\cal H}$ we have a (projective) isometric representation
of some space-time symmetry group
$g \mapsto U_{g}$
(say the Poincar\'e group), then by restriction to time
translation one can get
$U_{t}$.
To obtain
$\Phi({\bf x})$
the natural idea would be to search for a localization
observable in ${\cal H}$; this could be, for instance, the
Newton-Wigner-Wightman position operator [4]. If
$f_{\bf x}$
is a state in ${\cal H}$ describing a state localized in ${\bf
x}$, then a natural choics for
$\Phi({\bf x})$
is
$\Phi(f_{\bf x})$.
Finally, having
$\Phi(x)$
at our disposal, one must check if the couple
$(\Phi(x),U_{g})$
are linked by a relation of the type (1.1). This program,
although rather natural, had not been pursued in the
litterature. Instead it is favored another approach which we
call the {\it Weinberg anszatz}.

\subsection*{B The Weinberg Anszatz}

We note that if ${\cal H}$ is the Hilbert space carrying a
projective isometric irreducible representation of the
Poincar\'e group (or of one of its subgroups) then this space is
of the form:
$$
{\cal H} \subseteq L^{2}(\xm,\mas,{\bf C}^{N})
$$
where
$N = 1,2,...,\infty$.
In the cannonical formalism we have equality and in the fiber bundle
formalism we have a strict inclusion. We can identify then
${\cal H}^{(n)}$
with some subset of the Hilbert space of Borel maps
$\Phi^{(n)}_{k_{1},...,k_{n}} : (\xm)^{\times n} \rightarrow {\bf C}$
such that
$$
\sum_{k}\int_{(\xm)^{\times n}} \vert
\Phi^{(n)}_{k_{1},...,k_{n}}(p_{1},...,p_{n})\vert^{2}
\mas(p_{1})\cdots\mas(p_{n}) <\infty
$$

For the Fock space one must add:
$$
\Phi^{(n)}_{k_{\sigma(1)},...,k_{\sigma(n)}} = \pm
\Phi^{(n)}_{k_{1},...,k_{n}}
$$
(for the Bose case one has the sign $+$ and for the Fermi case
$(-1)^{\vert\sigma\vert}$).
The vector space
${\cal H}^{(n)}$
becomes a Hilbert space under the obvious scalar product.

One can introduce in $\foc$ the annihilation operators of fixed
momentum
$a_{k}(p)~~(k = 1,...,N;~p \in \xm)$
according to:
$$
\left(a_{k_{0}}(p_{0})\Phi)\right)^{(n)}_{k_{1},...,k_{n}} =
\sqrt{n+1}~\Phi^{(n+1)}_{k_{0},...,k_{n}}(p_{0},...,p_{n})\eqno(4.15)
$$
and also the creation operators of fixed momentum:
$$
a^{*}_{k}(p) = a_{k}(p)^{*}\eqno(4.16)
$$
interpreted in the weak sense. We note that
$a_{k}(p)$
is obtained from
$a_{\pm}(f)$
taking
$f_{k} = \sqrt{2p_{0}} \delta_{kk_{0}} \delta({\bf p} - {\bf
p_{0}})$.

Then, one makes for the free field
$\Phi(x)$
the following anszatz [10]:
$$
\Phi_{k}(x) = {1\over (2\pi)^{d-1\over 2}} \int_{\xm}
\mas \left[ e^{ip\cdot x} A_{kk'}(p) a_{k'}(p) +
e^{-ip\cdot x} B_{kk'}(p) a^{*}_{k'}(p) \right]\eqno(4.17)
$$
(summation over the dummy indices
$k,k',...$
is understood) and tries to find
$A_{kk'}$
and
$B_{kk'}$
such that (1.1) is fulfilled for some representation of $H$
living in
${\bf C}^{N}$.

For this one first notes that the representation of the
Poincar\'e group acting in $\foc$ is:
$$
{\cal U}_{h,a} \equiv \Gamma(U_{h,a})\eqno(4.18)
$$
(see (4.12)). Next, from the definition of the creation and
annihilation operators one easily finds out:
$$
{\cal U}_{h,a} a_{\pm}(f) {\cal U}_{h,a}^{-1} =
a_{\pm}(U_{h,a}f).\eqno(4.19)
$$

Now, it is easy to get the transformation properties of
$a_{k}(p)$.
For instance, in the cannonical formalism(see (2.3)) one gets:
$$
{\cal U}_{h,a} a_{k}(f) {\cal U}_{h,a}^{-1} =
e^{-ia\cdot\delta(h)p} \varphi^{\pi}_{kk'}(h^{-1},\delta(h)^{-1}\cdot p)
a_{k'}(\delta(h)\cdot p)\eqno(4.20)
$$
where
$\varphi^{\pi}_{kk'}$
is a cocyle associated to the representation
$\pi$
of the "little group" associated with the orbit $\xm$.

It is straightforward to prove that (4.17) will have a
transformation law of the type (1.1) only if the cocyles
$\varphi^{\pi}_{kk'}$
and
$\overline{\varphi^{\pi}_{kk'}}$
are cohomologous. According to [4], this is equivalent to the
statement that the representation
$\pi$
is self-contragrdient i.e.
$\pi$
and
$\overline{\pi}$
are unitary equivalent representations. Note that this condition
is only necessary.

In 1+3 dimensions one obtains the results of Weinberg. Let us
consider the proper orthochronous Poincar\'e group $\op$; for
systems of positive mass $m$ and spin $s$,
$\pi = D^{(s)}$
(= the representation of weight $s$ of $SU(2)$).
As
$D^{(s)}$
it is known to be self-contragradient one stands a chance to
obtain (1.1). That this is true, one can see from [10]. For
systems of zero mass and helicity
$n/2~~(n \in {\bf Z})$
the answer is negative: the representation
$\pi_{n}$
of
$\widetilde{SE(2)}$
inducing this case verifies
$\overline{\pi_{n}} \sim \pi_{-n}$.
So, to obtain a relativistic law of the type (1.1) one has to combine
$\pi_{n}$
and
$\pi_{-n}$;
for instance one could take
$\pi = \pi_{n} \oplus \pi_{-n}$.
According to [9] this representation of $\op$ admits an unique
extension to a representation of the orthochronous Poincar\'e
group, so to solve in an affirmative way Weinberg problem we
just might have started directly with this group, or with the
whole Poincar\'e group ${\cal P}$.

This observation shows that the existence of a convenient field
description of relativistic one-particle systems is somehow
connected with the consideration of inversions.

\subsection*{C Free Fields in 1+2 Dimensions}

Suppose that the one-particle system is a projective unitary
irreducible representation of $\op$ in 1+2 dimensions.
For positive mass these representations are induced by
$\pi^{(s)}$
(see (2.24)). It is clear that
$\overline{\pi^{(s)}} = \pi^{(-s)}$
so
$\pi^{(s)}$
is self-contragradient {\it iff}
$s = 0$.
So only in this case one can hope to construct a free field. If
we consider representations of the orthochronous Poincar\'e
group or of the whole Poincar\'e group ${\cal P}$, then the
situation is considerable improved. For the case of the
orthochronous Poincar\'e group,
$\tilde{D}^{(+)}$
and
$\tilde{D}^{(s)}$
are self-contragradient and for the case of ${\cal P}$ and
$\epsilon_{T} = 1$
the representations
$\tilde{\tilde{D}}^{(+)}$
and
$\tilde{\tilde{D}}^{(s)}$
are also self-contragradient. This properties follow easily form
(2.17).

It is only in this cases that we will prove the existence of a
convenient free field.

(i) First we consider the case of spin
$s = 0$.
The analysis is completely similar with the analysis in 1+3
dimensions. We have
${\cal H} = L^{2}(\xmp,\masp)$
and the represntation
$U_{h,a}$
is given by:
$$
\left(U_{h,a} f\right)(p) = e^{ia\cdot p} f(\delta(h)^{-1}\cdot p)\eqno(4.21a)
$$
$$
\left( S_{P} f\right)(p) = f(I_{P}\cdot p)\eqno(4.21b)
$$
$$
\left( S_{T} f\right)(p) = \overline{f(I_{S}\cdot p)}\eqno(4.21c)
$$

Then
${\cal H}^{(n)}$
is formed by Borel functions
$\Phi^{(n)} : (\xm)^{\times n} \rightarrow {\bf C}$
verifying
$$
\int_{(\xmp)^{\times n}} \masp (p_{1})\cdots\masp (p_{n})
\vert\Phi^{(n)}(p_{1},...,p_{n})\vert^{2} <\infty.
$$

To obtaint the Fock n-particle space one must take in the Bose case
$\Phi^{(n)}$
to be completely symmetric and in the Fermi case completely
antisymmetric.
The annihilation and creation operators are:
$$
\left(a(p_{0})\Phi\right)^{(n)}(p_{1},...,p_{n}) = \sqrt{n+1}~~
\Phi^{(n+1)}(p_{0},...,p_{n})\eqno(4.22a)
$$
$$
a^{*}(p) = a(p)^{*}.\eqno(4.22b)
$$

The free field is:
$$
\phi(x) = {1\over (2\pi)^{3/2}} \int_{\xmp} \masp \left[
e^{-ip\cdot x} a(p) + e^{ip\cdot x} a^{*}(p) \right].\eqno(4.23)
$$

Then one can verify the relativistic transformation laws:
$$
{\cal U}_{h,a} \phi(x) {\cal U}_{h,a}^{-1} = \phi(\delta(h)\cdot x+a)
\eqno(4.24a)
$$
$$
{\cal U}_{I_{P}} \phi(x) {\cal U}_{I_{P}}^{-1} = \phi(I_{P}\cdot x)\eqno(4.24b)
$$
$$
{\cal U}_{I_{T}} \phi(x) {\cal U}_{I_{T}}^{-1} = \phi(I_{T}\cdot x)\eqno(4.24c)
$$
i.e. something of the type (1.1).

Moreover, if one adopts Bose statistics the one has microcausality:
$$
[\phi(x),\phi(x')]_{-} = \Delta(x-x')\eqno(4.25)
$$
where the Pauli-Jordan distribution is:
$$
\Delta(x) = {m^{2}\over\pi} \theta(x^{2}) \varepsilon(x^{0})
{sin(m\sqrt{x^{2}})\over m\sqrt{x^{2}}} - m\delta(\vert{\bf
x}\vert-x^{0}) {x^{0}\over \vert{\bf x}\vert^{2}}\eqno(4.26).
$$

So,
$\Delta(x) = 0$
for
$x^{2} < 0$.

(ii) For particles of spin
$s > 0$
we take:
${\cal H} =  \tilde{\cal H}^{+,s}_{m}$
i.e. Borel functions
$f_{k\epsilon}: \xm \rightarrow {\bf C}$
verifying
$$
\int_{\xmp} \masp
\sum_{k\epsilon}\vert f_{k\epsilon}(p)\vert^{2} < \infty
$$
and the Dirac-like equation:
$$
\left( p\cdot H^{(\epsilon)} -\epsilon sm/2\right)_{kk'}
f_{k'\epsilon}(p) = 0.
$$

The representation of the full Poincar\'e group is:
$$
\left(U_{h,a} f\right)_{k\pm}(p) = e^{ia\cdot p}
D^{(s/2,\pm)}_{kk'}(h^{-1}) f_{k'\pm}(\delta(h)^{-1}\cdot p)\eqno(4.27a)
$$
$$
\left(S_{P} f\right)_{k\pm}(p) = f_{k\mp}(I_{P}\cdot p)\eqno(4.27b)
$$
$$
\left( S_{T} f\right)_{k\pm}(p) = (-1)^{k}
\overline{f_{k\mp}(I_{S}\cdot p)}.\eqno(4.27c)
$$

We remind the fact that the index $k$ follows the basis
$g_{k}$
from $F$ (see (2.23)).

Then
${\cal H}^{(n)}$
is formed by Borel functions
$\Phi^{(n)}_{k_{1}\epsilon_{1},...,k_{n}\epsilon_{n}} : (\xmp)^{\times n}
\rightarrow {\bf C}$
verifying
$$
\int_{(\xmp)^{\times n}} \masp (p_{1})\cdots\masp (p_{n})
\sum_{k_{i}\epsilon_{i}}\vert\Phi^{(n)}_{k_{1}\epsilon_{1},...
k_{n}\epsilon_{n}}(p_{1},...,p_{n})\vert^{2} <\infty.
$$
and also the Dirac-like equations:
$$
\left( p_{i}\cdot H^{(\epsilon_{i})} -\epsilon_{i} s/2 m\right)_{kk'}
\Phi_{k_{1}\epsilon_{1},...,k_{n}\epsilon_{n}}(p_{1},...,p_{n}) = 0.
$$

We have for the annihilation and creation operators the
following expressions:
$$
\left( a_{k_{0}}(p_{0},\epsilon_{0})\Phi\right)^{(n)}_{
k_{1}\epsilon_{1},...,k_{n}\epsilon_{n}}(p_{1},...,p_{n}) =
{}~\sqrt{n+1} \Phi^{(n+1)}_{
k_{0}\epsilon_{0},...,k_{n}\epsilon_{n}}(p_{0},...,p_{n})\eqno(4.28a)
$$
$$
a_{k}^{*}(p,\epsilon) = a_{k}(p,\epsilon)^{*}\eqno(4.28b)
$$

A convenient definition for the field operators is:
$$
\phi_{k\epsilon}(x) = {1\over (2\pi)^{3/2}} \int_{\xmp} \masp \left[
e^{-ip\cdot x} a_{k}(p,\epsilon) + e^{ip\cdot x} a_{k}^{*}(p,-\epsilon)
\right].\eqno(4.29)
$$
so it is clear why we did need the doubling of the representations.

Then one can verify the relativistic transformation laws:
$$
{\cal U}_{h,a} \phi_{k\pm}(x) {\cal U}_{h,a}^{-1} =
D^{(s/2,\pm)}(h^{-1}) \phi_{k'\pm}(\delta(h)\cdot x+a)\eqno(4.30a)
$$
$$
{\cal U}_{I_{P}} \phi_{k\pm}(x) {\cal U}_{I_{P}}^{-1} =
\phi_{k\mp}(I_{P}\cdot x)\eqno(4.30b)
$$
$$
{\cal U}_{I_{T}} \phi_{k\pm}(x) {\cal U}_{I_{T}}^{-1} =
\phi_{k\mp}(I_{T}\cdot x)\eqno(4.30c)
$$

Again, one can see that microcausality is compatible only with
Bose statistics. We close this section with some remarks.

{\bf Remark 3} For every value of
$s \in \R\cup\{0\}$
the Klein-Gordon equation
$$
(\partial^{2} + m^{2}) \phi(x) = 0\eqno(4.31)
$$
is verified.

{\bf Remark 4} It is easy to see that one has more general
expressions of the type (4.29) verifying (4.30), (4.31) and the
microcausality.  In 1+3 dimensions one usually eliminates this
freedom imposing additional equations of Dirac type on the field
operator. This procedure does not work in 1+2 dimensions.

{\bf Remark 5} One can see why the construction from [7] is
compatible with a convenient relativistic transformation law.
Indeed in [7] one takes
$s \in \N/2$
and
$\pi = \pi^{(-s)} \oplus \pi^{(-s+1)}\oplus\cdots\oplus \pi^{(s)}$
which is self-contragradient. Moreover,
$W^{\xmp,\pi}$
admits an extension to ${\cal P}$ which can be used for the
construction of the one-particle Hilbert space (it will not be
irreducible). This idea is also used in [11] for zero-mass particles
in 1+3 dimensions.

{\bf Remark 6} It is natural to define the charge conjugation
operator as follows:

- for
$s = 0$:
$$
({\cal U}_{C}\Phi)^{(n)}(p_{1},...,p_{n}) =
\overline{\Phi^{(n)}(p_{1},...,p_{n})}\eqno(4.32)
$$

-for
$s > 0$
$$
({\cal U}_{C}\Phi)^{(n)}_{k_{1}\epsilon_{1},...,k_{n}\epsilon_{n}}
(p_{1},...,p_{n}) =
\overline{\Phi^{(n)}_{k_{1}-\epsilon_{1},...,k_{n}-\epsilon_{n}}
(p_{1},...,p_{n})}\eqno(4.33)
$$

Then one has in the first case:
$$
{\cal U}_{C} \phi(x) {\cal U}_{C}^{-1} = \phi(x)^{*}\eqno(4.34)
$$
and in the second case:
$$
{\cal U}_{C} \phi_{k\epsilon}(x) {\cal U}_{C}^{-1} =
\phi(x)_{k-\epsilon}^{*}\eqno(4.35)
$$

{\bf Remark 7} One can make formally
$m \rightarrow 0$
in all the formul\ae~above.

\section{ PCT, spin and statistics and all that in 1+2 dimensions}

Let us see if the main theorems of axiomatic field theory can be
obtained in 1+2 dimensions. From [1] it folows that we have to
check the validity of some steps from the 1+3 dimensional
argument. We will suppose that the Wightman field is defined by
the framework from section 1, where in (1.1) one must take
$T = 1$
(for spin
$s = 0$)
or
$T = D^{(s/2,+)} \oplus D^{(s/2,-)}$
(for spin
$s > 0$).

\subsection*{A The Complex Lorentz Group $L({\bf C})$}

This group is defined in 1+2 dimensions in strict analogy with
the 1+3 dimensional case. As in [1], p.13 one shows that
$L({\bf C})$
has two connected components
$L_{\pm}({\bf C})$
with determinant
$\pm 1$
respectively. Indeed, the continous path:
$$
\Lambda(t) \equiv \left(\matrix{ch(it) & 0 & sh(it)\cr
0 & 1 & 0\cr sh(it) & 0 & ch(it)\cr}\right)\eqno(5.1)
$$
links
$\Lambda(0) = {\bf 1}$
with
$\Lambda(\pi) = I_{PT}$.

\subsection*{ B The Relativistic Covariance of Wightman Functions}

In analogy with (2.84) from [1] one gets from (1.1) that we have
for
$(\zeta_{1},...,\zeta_{n})$
in the tube
${\cal T}_{n}$
the following identities for the Wightman functions:

- for
$s = 0$:
$$
{\cal W}(\delta(h)\cdot \zeta_{1},...,\delta(h)\cdot \zeta_{n}) =
{\cal W}(\zeta_{1},...,\zeta_{n})\eqno(5.2)
$$

- for
$s > 0$:
$$
{\cal W}_{k_{1}\epsilon_{1},...,k_{n}\epsilon_{n}}
(\delta(h)\cdot \zeta_{1},...,\delta(h)\cdot \zeta_{n}) =
\Pi_{i=1}^{n} D^{(s/2,\epsilon_{i})}_{k_{i}k_{i}'}(h)
{\cal
W}_{k'_{1}\epsilon_{1},...,k'_{n}\epsilon_{n}}(\zeta_{1},...,\zeta_{n})
\eqno(5.3)
$$

If we take
$h = (\pi,0)$
one easily obtains that
${\cal W}_{k_{1}\epsilon_{1},...,k_{n}\epsilon_{n}}$
can be non-zero only if
$2s\sum\epsilon_{i} \in {\bf Z}$.

\subsection*{C Prolongation to the Extended Tube
${\cal T}_{n}'$}

(i) We first consider the case
$s = 0$.
One would like to obtain from (5.2) that:
$$
{\cal W}(\Lambda\cdot \zeta_{1},...,\Lambda\cdot \zeta_{n}) =
{\cal W}(\zeta_{1},...,\zeta_{n})\eqno(5.4)
$$
for any
$(\zeta_{1},...,\zeta_{n}) \in {\cal T}_{n}'$
and for any
$\Lambda \in L_{+}({\bf C})$.

For this one needs to check if the lemma appearing in [1] at p.
66 is still true, i.e. if for any
$(\zeta_{1},...,\zeta_{n}) \in {\cal T}_{n}$
and
$\Lambda \in L_{+}({\bf C})$
such that:
$(\Lambda\cdot \zeta_{1},...,\Lambda\cdot \zeta_{n}) \in {\cal T}_{n}$
there exists a continous path
$\Lambda(t) \in L_{+}({\bf C})$
with
$\Lambda(0) = {\bf 1}$,
$\Lambda(1) = \Lambda$
and
$(\Lambda(t)\cdot \zeta_{1},...,\Lambda(t)\cdot \zeta_{n}) \in {\cal
T}_{n}~~\forall t \in [0,1]$.

To prove this one first proves that every element
$\Lambda \in L_{+}({\bf C})$
can be written as a product of elements of the type:
$$
\Lambda_{1} \equiv \left(\matrix{cos(\phi) & 0 & isin(\phi)\cr
0 & 1 & 0\cr isin(\phi) & 0 & cos(\phi)\cr}\right)\eqno(5.5)
$$
$$
\Lambda_{2} \equiv \left(\matrix{cos(\phi) & i sin(\phi) & 0\cr
i sin(\phi) & cos(\phi) & 0\cr 0 & 0 & 1\cr}\right)\eqno(5.6)
$$
$$
\Lambda_{3} \equiv \left(\matrix{ch(\chi) & 0 & sh(\chi)\cr
0 & 1 & 0\cr sh(\chi) & 0 & ch(\chi)}\right)\eqno(5.7)
$$

It is sufficient to check the lemma for
$\Lambda_{1}, \Lambda_{2}, \Lambda_{3}$
separately. We do this, for illustration, on the case
$\Lambda_{1}$.
First we can easily arrange that
$\phi \in [0,\pi)$.
It is obvious that one should try to work with the path:
$$
\Lambda_{1}(t) \equiv \left(\matrix{cos(t\phi) & 0 & isin(t\phi)\cr
0 & 1 & 0\cr isin(t\phi) & 0 & cos(t\phi)\cr}\right)\eqno(5.8)
$$

If
$z \in {\bf C}^{3}$
and
$z(t) \equiv \Lambda(t)\cdot z$,
then
$z(t) = \xi(t) - i\eta(t)$
with
$\xi(t), \eta(t) \in \R^{3}$.
One must prove that
$\eta(t) \in V_{+} \equiv \{ x \in \R^{3}\vert x^{0} > \vert{\bf x}\vert\}$.
In our case:
$$
\eta(t) = (y^{0} cos(t\phi) + x^{2} sin(t\phi), 0,
y^{2} cos(t\phi) - x^{0} sin(t\phi))\eqno(5.9)
$$
where:
$z = x - iy~~(x, y \in \R^{3})$.

One uses the fact that
$\eta \in V_{+}$
{\it iff}
$n\cdot\eta > 0$
for every
$n \in C_{+} \equiv \{ x \in \R^{3}\vert x^{0} = \vert{\bf x}\vert\}$.
In our case:
$$
sin(\phi) n\cdot\eta(t) = sin((1-t)\phi) n\cdot\eta(0) +
sin(t\phi) n\cdot\eta(1).\eqno(5.10)
$$

So,
$n\cdot\eta(t)$
is positive if
$n\cdot\eta(0)$
and
$n\cdot\eta(1)$
are, which is true by assumption.

The same argument works for
$\Lambda_{2}$
and
$\Lambda_{3}$
also. It follows that (5.4) is valid for every
$\Lambda \in L_{+}({\bf C})$
and any
$(\zeta_{1},...,\zeta_{n})$
in the extended tube
${\cal T}_{n}'$.
In particular, for
$\Lambda = I_{PT}$
we get:
$$
{\cal W}(I_{PT}\cdot\zeta_{1},...,I_{PT}\cdot\zeta_{n}) =
{\cal W}(\zeta_{1},...,\zeta_{n})\eqno(5.11)
$$
for every
$(\zeta_{1},...,\zeta_{n}) \in {\cal T}_{n}'$.

(ii) Now we may wonder if the same argument works for
$s > 0$
i.e. (5.3) can be extended to
${\cal T}_{n}'$.
For this one must look for a complex extension of the representations
$D^{(s/2,\epsilon)}$.
In 1+3 dimensions one can do the trick using the universal
covering group of
$L_{+}({\bf C})$.
This is
$SL(2,{\bf C}) \times SL(2,{\bf C})$
which contains in a natural way the universal covering group of
$\op$ i.e.
$SL(2,{\bf C})$
(as the diagonal subgroup).

In 1+2 dimensions, the situation is completely different. The
universal covering group of
$L_{+}({\bf C})$
is
$SL(2,{\bf C})$.
Indeed, the covering homomorphism is given by:
$$
[\delta(A)z] = A [z] A^{t}\eqno(5.12)
$$
where for every
$z \in {\bf C}^{3}$
we have denoted:
$$
[z] = \left(\matrix{z^{0}+z^{1} & z^{2}\cr z^{2} &
z^{0}-z^{1}\cr}\right).\eqno(5.13)
$$

Now it is easy to see that $\ul$ cannot be imbeded in
$SL(2,{\bf C})$
(mainly for topological reasons). So, it seems that (5.3) cannot
be extended as at (i); it follows that we do not have a relation
of the type (5.4) (or (5.11)) for
$s > 0$.

\subsection*{ D PCT, Spin and Statistics for $s = 0$}

We try to work out thm. 4-6 from [1] in the case
$s = 0$.
One has to combine the condition of weak local commutativity (WLC):
$$
{\cal W}(\zeta_{1},...,\zeta_{n-1}) =
{\cal W}(-\zeta_{n-1},...,-\zeta_{1})\eqno(5.14)
$$
with (5.11) (with
$n \rightarrow n-1$)
to obtain
$$
{\cal W}(\zeta_{1},...,\zeta_{n-1}) =
{\cal W}(-I_{PT}\cdot\zeta_{n-1},...,-I_{PT}\cdot\zeta_{1})\eqno(5.15)
$$

This implies the PCT condition on the vacuum expectation values:
$$
(\Phi_{0},\phi(x_{1}),...,\phi(x_{n})\Phi_{0}) =
(\Phi_{0},\phi(I_{PT}\cdot x_{n}),...,\phi(I_{PT}\cdot x_{1})\Phi_{0}).
\eqno(5.16)
$$

As usual (5.16) implies the existence of the PCT operator
$\Theta$
which verifies:
$$
\Theta \phi(x) \Theta^{-1} = \phi(I_{PT}\cdot x)^{*}.\eqno(5.17)
$$

The argument of thm. 4-9 of [1] can be generalized along similar
lines.

\end{document}